\documentclass[journal]{IEEEtran}

\usepackage{cite}
\usepackage{graphicx}
\usepackage{amsmath}
\usepackage{amsfonts}
\usepackage{balance}
\usepackage{xcolor}
\usepackage{multirow}
\usepackage{booktabs}
\usepackage[caption=false]{subfig}
\usepackage{flushend}
\usepackage{tikz}

\newtheorem{theorem}{Theorem}

\newtheorem{lemma}{Lemma}

\newtheorem{corollary}{Corollary}

\newtheorem{proposition}{Proposition}

\newtheorem{conjecture}{Conjecture}

\newtheorem{sketch}{Sketch of Proof}

\newcommand\submittedtext{%
  \footnotesize This work has been submitted to the IEEE for possible publication. Copyright may be transferred without notice, after which this version may no longer be accessible.}

\newcommand\submittednotice{%
\begin{tikzpicture}[remember picture,overlay]
\node[anchor=north,yshift=-10pt] at (current page.north) {\fbox{\parbox{\dimexpr0.65\textwidth-\fboxsep-\fboxrule\relax}{\submittedtext}}};
\end{tikzpicture}%
}

\DeclareMathOperator*{\argmax}{argmax}

\belowdisplayskip=\abovedisplayskip
\begin{document}

\title{Deep Learning-Based Computer Vision for Beam Selection and Proactive Blockage Prediction}

\author{Sachira Karunasena,~\IEEEmembership{Student Member,~IEEE,}
        Erfan Khordad,~\IEEEmembership{Member,~IEEE,}
        Tom Drummond,~\IEEEmembership{Member,~IEEE,}
        and~Rajitha Senanayake,~\IEEEmembership{Member,~IEEE}
\thanks{S.~Karunasena, E.~Khordad and R.~Senanayake are with the Department of Electrical and Electronic Engineering, University of Melbourne, VIC 3010, Australia (e-mail: s.karunasena@unimelb.edu.au, erfan.khordad@unimelb.edu.au, rajitha.senanayake@unimelb.edu.au).}
\thanks{T.~Drummond is with the School of Computing and Information Systems at University of Melbourne, VIC 3010, Australia (e-mail: tom.drummond@unimelb.edu.au).}}


\maketitle

\renewcommand\fbox{\fcolorbox{red}{white}}
\setlength{\fboxrule}{2pt} 
\submittednotice

\begin{abstract}
Millimeter-wave communication faces two critical challenges: propagation losses requiring costly narrow-beam alignment, and penetration losses causing link failures from blocked line-of-sight paths. We address propagation loss through a novel vision-aided beam selection framework that integrates RGB imagery with received power profiles for efficient transmitter identification and beam prediction. This framework achieves 98.96\% top-5 beam prediction accuracy, surpassing current state-of-the-art methods by at least 6\% across all metrics. We address penetration loss through a proactive blockage prediction framework using a modified object tracker with weighted centroid-based depth estimation. This represents the first analysis of simultaneous non-uniform mobility of both transmitters and obstacles. Evaluated on completely unseen data, this framework achieves over 98\% accuracy in predicting blockages up to three frames ahead, establishing strong performance benchmarks. 
\end{abstract}


\IEEEpeerreviewmaketitle

\section{Introduction}
The escalating demand for ultra-high data rates in modern wireless networks has necessitated the exploration of millimeter-wave (mmWave) and sub-terahertz (sub-THz) spectrum bands for next-generation communication systems. Operating at these elevated frequencies enables unprecedented throughput capabilities essential for emerging applications \cite{rappaport2019}. However, high-frequency communication faces two key constraints that limit its practical deployment. The short wavelengths inherent to mmWave signals result in substantial propagation losses over distance \cite{li2020}, while simultaneously exhibiting extreme susceptibility to blockages from environmental obstacles \cite{andrews2016}. 
These two challenges undermine mmWave link reliability, particularly in dynamic environments where there are mobile transmitters (TX) and obstacles \cite{khan2023, alkhateeb2018}.

In general, to compensate for the severe propagation losses inherent to mmWave frequencies, large-scale antenna arrays that can generate highly directional beams are employed. As the array aperture increases, these beams become correspondingly narrower to achieve the required directivity. 
Such narrow-beam operation necessitates a substantially larger set of beam patterns to ensure adequate spatial coverage, thereby expanding the size of the underlying beam codebook and requiring efficient methods to identify the optimal beam directions for reliable communication.
Beam selection in mmWave systems has progressed through several evolutionary stages. The foundational approach, exhaustive beam sweeping (EBS) \cite{ebs}, systematically evaluates predetermined beam codebooks at both the TX and receiver (RX). This method identifies the optimal beam pair by measuring received signal power across all candidate beams. To reduce the computational burden of exhaustive search, more efficient techniques have been developed, including hierarchical tree search algorithms \cite{xiao2016hierarchical, hur2013millimeter} and multi-disciplinary codebook designs \cite{yan2019wideband, lin2016terahertz}.
Although straightforward and robust, the training overhead of EBS scales linearly with the codebook size. Consequently, the large codebooks associated with narrow-beam arrays render exhaustive search inefficient in terms of both computational load and training latency. 
These limitations are particularly pronounced in latency-sensitive applications and highly dynamic environments, where rapid and reliable beam adaptation is essential.

Recent work has explored Machine Learning (ML) based solutions to the beam selection problem associated with propagation loss in mmWave systems, with methods operating both with \cite{nie2023, imran23, kim2024, klautau2019lidar, morais2023position} and without \cite{cousik22, ma2021deep, qi2020deep, alrabeiah2020deep} auxiliary sensory information. These approaches leverage data sources such as Global Positioning System (GPS) coordinates \cite{morais2023position}, Light Detection and Ranging (LiDAR) point clouds \cite{klautau2019lidar}, and Red, Green, Blue (RGB) imagery \cite{nie2023, imran23, kim2024} to reduce the computational burden of EBS while maintaining competitive prediction accuracy. 

Vision-based methods in particular \cite{nie2023, kim2024} rely solely on RGB features for TX identification and beam prediction, omitting mmWave channel characteristics. Such purely vision-driven strategies may face generalization challenges when TX appearance varies across deployments.
Likewise, the approach in \cite{imran23} depends exclusively on RGB imagery and semantic segmentation, foregoing potential gains achievable through the inclusion of received power measurements.
In contrast, our framework explicitly integrates both images and mmWave signal information for enhanced TX identification, which leads to enhanced beam selection performance. 

Due to the high-frequency nature of mmWave signals, physical obstacles introduce substantial attenuation, often causing severe link degradation or complete disconnection when the line-of-sight (LoS) path between the base station (BS) and user equipment is obstructed \cite{alkhateeb2018}. 
A key strategy to address this vulnerability lies in proactive blockage prediction. By anticipating potential obstructions before they materialize, the network can execute preventive measures such as user handoff to alternative BS or frequency band switching from mmWave to more resilient sub-6 GHz carriers in co-deployed heterogeneous networks \cite{semiari2019, polese2017, petrov2017, ali2019early}. Accurate blockage prediction thus emerges as a critical capability for ensuring mmWave link reliability, enabling the system to maintain connectivity through preemptive adaptation rather than reactive recovery after link failure has already occurred 

Proactive blockage prediction has been explored through both classical signal processing techniques and ML-based approaches. Classical methods, although effective in controlled settings, are tailored to simplified mobility patterns, such as stationary users with linearly moving blockers or mobile users on straight trajectories with static obstacles \cite{alkhateeb2018, shunyao20222}.
Machine learning frameworks have emerged as more capable alternatives, particularly those incorporating auxiliary sensing modalities. Radar- and LiDAR-based systems \cite{demirhan2022radar, wu2022lidar} demonstrate the feasibility of early blockage detection through high-resolution environmental sensing. However, radar performance deteriorates in dense environments due to multipath and clutter, whereas LiDAR sensors, despite their accuracy, incur high deployment cost and offer limited range \cite{charan2022computer}. Vision-based approaches leveraging RGB imagery and Vision Transformers \cite{charan2022computer, ghassemi2025ViT} provide a cost-effective alternative with broad spatial coverage. Similar to classical methods, existing vision-based solutions have been evaluated only under simplified mobility conditions: either stationary users with linear blockage motion or mobile users traveling along linear paths with static obstacles \cite{charan2022computer, ghassemi2025generative, alrabeiah2020millimeter}.
Importantly, no prior work considers the realistic scenario in which both the user equipment and potential blockers exhibit simultaneous, non-uniform mobility, despite this representing the predominant operating condition in practical settings such as vehicular networks and dense urban deployments.

In this paper, we propose two frameworks for beam selection and blockage prediction to focus on challenges associated with propagation and penetration losses in mmWave communication. 
More specifically, for propagation loss, we develop a vision-aided beam selection architecture that fuses RGB imagery with mmWave received power profiles to identify TXs and predict beams jointly. Unlike prior work \cite{charan23} that relies on separate learning models, our unified Deep Learning (DL) framework reduces latency and computational overhead while achieving robust generalization by identifying TXs based on mmWave signal characteristics, regardless of their visual appearance.
Our beam prediction component further incorporates perspective-aware corrections, explicitly modeling vertical vanishing point distortions induced by camera geometry to ensure geometrically consistent beam direction estimation across deployments. 

For penetration loss, we introduce the first blockage prediction framework designed for environments where both the user equipment and multiple obstacles exhibit simultaneous, non-uniform mobility. 
Our methodology features two core innovations: First, a modified tracking algorithm that extrapolates user and obstacle trajectories to predict future LoS intersections, supported by a weighted centroid-based depth estimator. Second, a post-blockage recovery mechanism that rapidly re-identifies and re-associates the user equipment following blockage clearance, combining motion-based prediction with definitive DL-based TX re-identification. 
We evaluate both frameworks on the DeepSense 6G dataset \cite{deepsense6g}, the only publicly available real-world multimodal mmWave dataset.
Collectively, these contributions provide a comprehensive systematic analysis of realistic multi-agent mobility in proactive mmWave blockage prediction, addressing a key gap between existing research assumptions and practical deployment requirements.

Our prior conference paper \cite{ours} presented the foundational framework for TX identification and vision-aided beam prediction.
The novel contributions of this paper extend our work to penetration loss mitigation and include: 
\begin{itemize}
    
    \item \textbf{A novel TX identification module:} A  module which leverages mmWave power profiles to detect the TX in the BS field of view (Section~\ref{sec:TX_identi}).

    \item \textbf{Geometrically aware beam prediction system:} A module that incorporates vertical vanishing points to improve the accuracy of predicting top-$N$ beams under varied imaging scenarios (Section \ref{sec:beam_pred}).

    \item \textbf{Modified Object Tracking Algorithm:} A trajectory extrapolation framework capable of predicting future spatial positions of both user equipment and dynamic obstacles based on recent motion history (Section \ref{sec:TX_track_bl}).
    
    \item \textbf{Depth Estimation-Based Blockage Detection:} A weighted centroid-based depth analysis methodology that accurately determines relative distance ordering between the TX and potential obstructions to identify genuine blockage threats (Section \ref{sec:blockage}).
    
    \item The first systematic investigation of blockage prediction for scenarios with simultaneous non-uniform mobility of both user equipment and multiple dynamic blockers, establishing performance benchmarks for realistic urban deployment conditions.
    
    \item \textbf{Post-Blockage TX Re-identification and Recovery:} A dual-stage mechanism combining motion-based prediction with definitive DL-based TX re-identification to rapidly restore beam alignment and communication following blockage clearance. (Section \ref{sec:blockage})

    \item We present comprehensive numerical results comparing our frameworks against existing methodologies. Our propagation loss mitigation framework surpasses current state-of-the-art vision-aided beam selection methods \cite{nie2023, imran23} by at least 6\% across all evaluation metrics. Our penetration loss mitigation framework, being the first to address simultaneous dynamic mobility of both TX and obstacles establishes initial performance benchmarks for this previously unexplored scenario, achieving over 99\% blockage prediction accuracy.
\end{itemize}

\section{System Model}
\label{sec:sysModel}

\begin{figure}[t]
    \centering
    \includegraphics[width=0.48\textwidth]{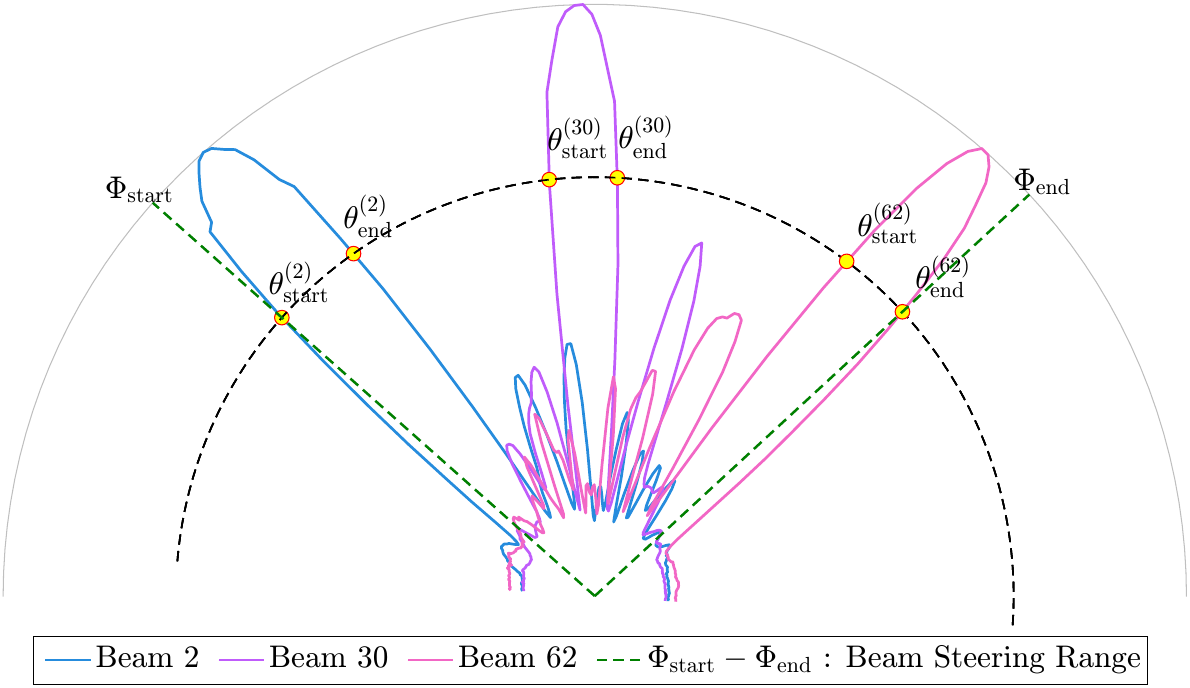}
    \caption{[$\theta^{(q)}_{\text{start}}, \theta^{(q)}_{\text{end}}]$ of selected 3 beams and the beam steering range of the BS, $[\Phi_{\text{start}}, \Phi_{\text{end}}]$ from the DeepSense 6G \cite{deepsense6g} codebook. These values are used to define the corresponding x-region boundaries as defined in \eqref{eqn:x_region}.}
    \label{fig:beam_pattern}
\end{figure}

We consider an Orthogonal Frequency Division Multiplexing (OFDM)-based mmWave communication system wherein the BS is equipped with an $M$-element uniform linear array and an RGB camera, while the mobile RX under consideration utilizes a single omnidirectional antenna. The system operates across $K$ OFDM subcarriers, and the BS employs an oversampled analog beamforming codebook denoted as $\mathcal{C}_r=\{\textbf{f}_q\}_{q=1}^Q$, where $Q$ designates the codebook cardinality and each beamforming vector satisfies $\textbf{f}_q \in \mathbb{C}^{M \times 1}$. Each $\textbf{f}_q$ corresponds to a narrow directional beam spanning a specific azimuth angular sector $[\theta^{(q)}_{\text{start}}, \theta^{(q)}_{\text{end}}]$ within the beam steering range of the BS $[\Phi_{\text{start}}, \Phi_{\text{end}}]$ as depicted in Fig. \ref{fig:beam_pattern}. 

For such a system, we can write the received signal on the $k$-th subcarrier at time $t$ as:
\begin{equation}
    y_k(t) = \textbf{h}_k^T(t)\textbf{f}_q(t)x_k(t) + v_k(t),
    \label{eqn:rx_signal}
\end{equation}
where $v_k(t)$ denotes complex additive white Gaussian noise at the receiver following $\mathcal{CN}(0, \sigma^2)$, $x_k(t)$ represents the transmitted symbol at time $t$, $\textbf{h}_k^T(t)$ represents the channel between the BS and RX over the $k$-th subcarrier, and $\textbf{f}_q(t)$ represents the applied beamforming vector from codebook $\mathcal{C}_r$ at time $t$. The beamforming objective is to identify the optimal beamforming vector $\textbf{f}^*(t)$ that maximizes the subcarrier-averaged received signal-to-noise ratio (SNR) over all $k \in [1, K]$ and all candidate vectors $\{\textbf{f}_q\}_{q=1}^Q$. Formally, $\textbf{f}^*(t)$ is obtained via:
\begin{equation}
    \textbf{f}^*(t) = \argmax_{\textbf{f}_q(t) \in \mathcal{C}_r} \frac{1}{K} \sum^{K}_{k=1} \Vert \textbf{h}_k^T(t)\textbf{f}_q(t) \Vert ^2 \mathsf{SNR},
    \label{eqn:beam_argmax}
\end{equation}
where $\mathsf{SNR} = \frac{\mathbb{E}[\Vert x_k(t) \Vert^2]}{\sigma^2}$ characterizes the transmit power-to-noise ratio. In subsequent sections, we develop a vision-aided deep learning framework to efficiently determine $\textbf{f}^*(t)$ without exhaustive codebook search. 

The channel $\textbf{h}_k(t)$ can be decomposed into LoS and non-line-of-sight (NLoS) propagation components:
\begin{equation}
    \textbf{h}_k(t) = \beta_{\text{LoS}}(t) \textbf{h}_k^{\text{LoS}}(t) + \beta_{\text{NLoS}}(t) \textbf{h}_k^{\text{NLoS}}(t),
    \label{eqn:channel_decomposition}
\end{equation}
where $\textbf{h}_k^{\text{LoS}}(t) \in \mathbb{C}^{M\times1}$ and $\textbf{h}_k^{\text{NLoS}}(t) \in \mathbb{C}^{M\times1}$ represent the LoS and NLoS channel vectors respectively, and $\beta_{\text{LoS}}(t), \beta_{\text{NLoS}}(t) \in \{0,1\}$ are binary indicator variables denoting the presence of each propagation path. These binary coefficients satisfy the constraint $\beta_{\text{LoS}}(t) + \beta_{\text{NLoS}}(t) = 1$. We formally define a blockage event at time $t$ as the condition:
\begin{equation}
    \text{Blockage}(t) = \begin{cases}
        1, & \text{if } \beta_{\text{LoS}}(t) = 0, \\
        0, & \text{otherwise},
    \end{cases}
    \label{eqn:blockage_definition}
\end{equation}
indicating complete obstruction of the direct propagation path between the BS and TX.

\section{Vision-aided Beam Prediction}
\label{sec:va-bp}
\begin{figure*}[t]
\vspace*{0.1in}
    \centering
    \includegraphics[width=1\textwidth]{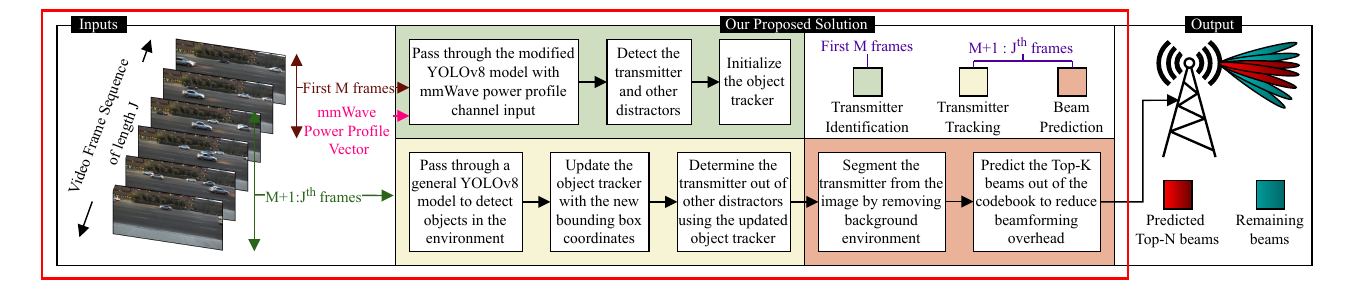}
    \caption{Proposed end-to-end Vision-aided Beamforming approach (Red box). The first $\mathbf{M}$ frames of the video sequence will be used for TX identification. The remaining $\mathbf{M+1:J^{th}}$ frame is used to track the detected TX and predict the top-$N$ beams for each frame. Proposed framework consists of three stages, 1. TX Identification (highlighted in green), 2. TX Tracking (highlighted in yellow) and 3. Beam Prediction (highlighted in orange).}
    \label{fig:framework1}
\end{figure*}

This section presents our proposed data-driven end-to-end beam selection methodology that uses RGB images and mmWave received signal power profiles as inputs. The framework consists of three key sections as illustrated in Fig. \ref{fig:framework1}.
\begin{enumerate}
    \item \textbf{Transmitter Identification (Section \ref{sec:TX_identi})}: Precisely locating the TX amid multiple interfering objects within the surrounding environment.
    \item \textbf{Transmitter Tracking (Section \ref{sec:TX_track})}: Continuously monitoring the identified TX across consecutive frames while it remains within the beam steering range of the BS.
    \item \textbf{Beam Prediction (Section \ref{sec:beam_pred})}: Estimating the top-N candidate beams for serving the monitored TX.
\end{enumerate}

\subsection{Transmitter Identification}
\label{sec:TX_identi}

Accurate TX identification within the BS coverage area represents a fundamental challenge when multiple mobile objects are present. RGB imagery alone cannot distinguish the target TX from other mobile entities, creating inherent ambiguity in multi-object scenarios. To address this limitation, we integrate the mmWave power profile as an additional input channel. This modified image with the additional mmWave power profile channel is then fed into our object detection framework to leverage the directional characteristics of the mmWave power propagation to identify the TX among multiple detected objects.

Our proposed framework integrates mmWave power profiles directly into the object detection process through a unified single-inference architecture, where the TX identification is achieved in one forward pass through the network.
This approach eliminates the computational overhead associated with cascaded ML models, as employed in \cite{charan23}, thereby achieving reduced latency and parameter complexity compared to sequential processing across multiple models.
Consequently, existing methods described in \cite{nie2023,kim2024} do not offer a generalized solution applicable across diverse TX types. 
Our proposed methodology addresses this fundamental limitation by systematically suppressing visual cues in the training process. 
This compels the model to perform TX identification based exclusively on mmWave power profile signatures that reflect the true physical relationship between signal propagation and TX location.
Our proposed integration methodology transforms the standard RGB image input into a modified three-channel configuration, where each channel encodes distinct data dimensions analogous to the conventional color channel decomposition. The three channels are as follows, and each is detailed subsequently:
\begin{itemize}
    \item \textbf{Channel 1:} Visual feature representation (detailed in subsequent methodological variants).
    \item \textbf{Channel 2:} MmWave power profile data spatially aligned in the BS coverage area, where signal intensity maxima correspond to TX location/s.
    \item \textbf{Channel 3:} Zero-padding channel to maintain compatibility with pre-trained MS COCO weights \cite{cocodataset} for transfer learning optimization.
\end{itemize}

\begin{figure}[t]
    \centering
    \includegraphics[width=0.5\textwidth]{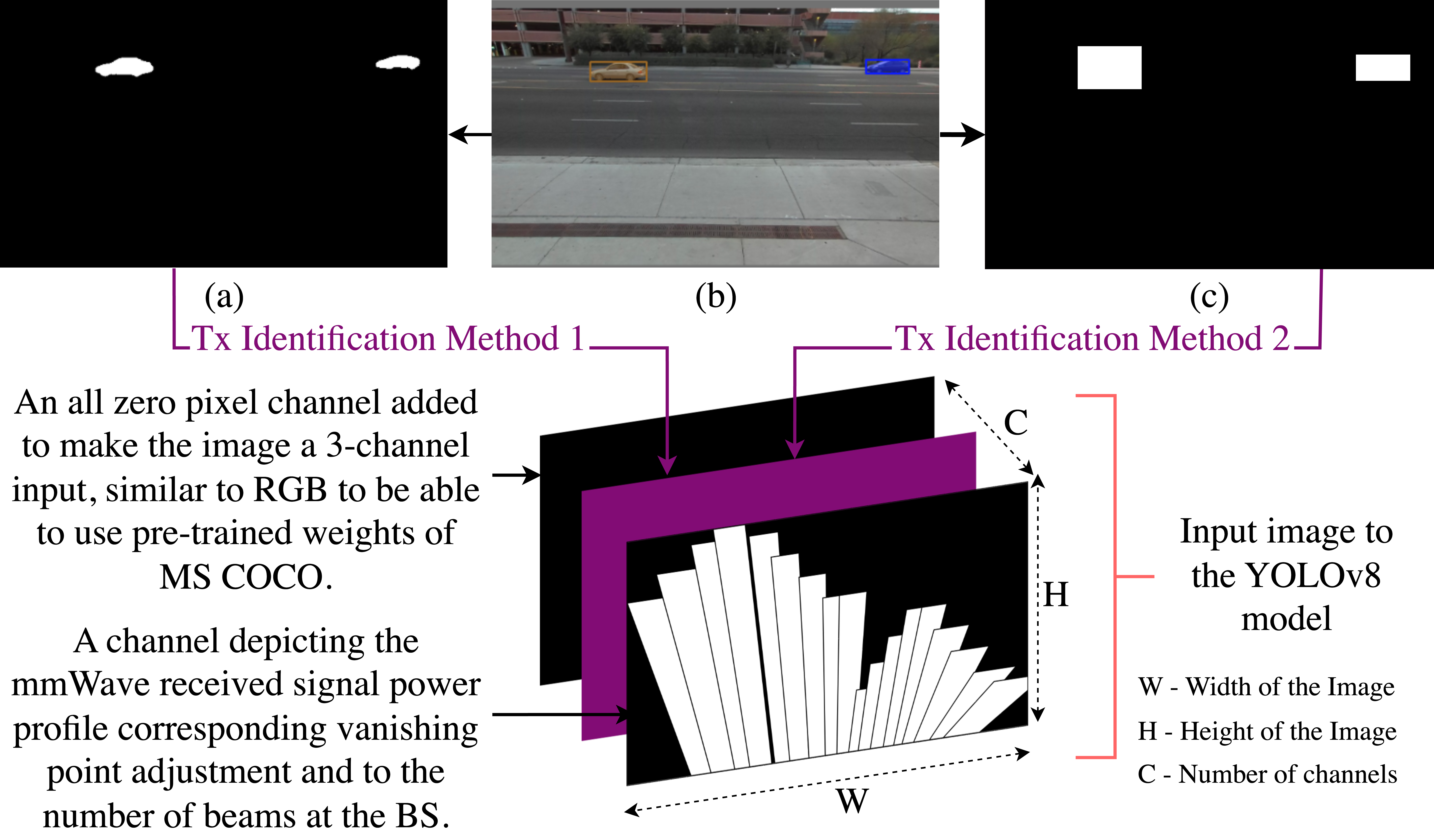}
    \caption{Structuring the input for the TX Identification methods. This input will be used as an input to the TX Identification stage of Fig. \ref{fig:framework1}. Sub-figures (a) and (c) represent the input channel according to the two TX identification methods while the sub-figure (b) represents the original image from the dataset.}
    \label{fig:tx_identi}
\end{figure}

\subsubsection{Structuring Channel 1: Visual Feature Representation}

To effectively integrate the mmWave power profile with visual information, we propose two distinct methodologies to structure the input to the YOLOv10 object detector \cite{yolov10}. These approaches are specifically designed to control the extent to which the model can rely on visual cues versus mmWave channel characteristics during TX identification. A critical challenge in the development dataset \cite{deepsense6g} is that the TX consistently corresponds to the same physical object throughout the dataset. This consistency introduces a potential confounding factor that a naively trained model might learn to identify the TX based on appearance specific features rather than the underlying mmWave propagation characteristics that we seek to leverage. To ensure that our model develops a genuine understanding of the relationship between mmWave power profiles and TX location, rather than simply memorizing visual patterns associated with a particular individual, we systematically ablate visual information from the training data.
Both methodologies begin with a common preprocessing step wherein background regions are segmented and removed from the image, retaining only objects of interest within the scene. Subsequently, we apply different levels of visual information suppression:

\begin{enumerate}
    \item \textbf{Method 1: Shape-Preserving:} All color information is removed while preserving geometric structure, leaving grayscale silhouettes that maintain the original shape and contour characteristics of each object. This eliminates color-based discriminative features, compelling the model to fuse geometric information with the mmWave power profile for TX identification. 
    Fig. \ref{fig:tx_identi}(a) is used to construct the three-channel input image shown therein in this method. 
    \item \textbf{Method 2: Complete Visual Ablation:} Both chromatic and geometric information are eliminated from all objects within the scene. Objects are replaced with uniform masks, such that no shape, texture, or color information remains. The model must rely exclusively on the mmWave beam power profile for TX detection and localization.
    Fig. \ref{fig:tx_identi}(c) is used to construct the three-channel input image shown therein in this method. 
    
\end{enumerate}

\subsubsection{Structuring Channel 2: mmWave Power Profile Mapping}

\begin{figure}[!t]
    \centering
    \includegraphics[width=0.48\textwidth]{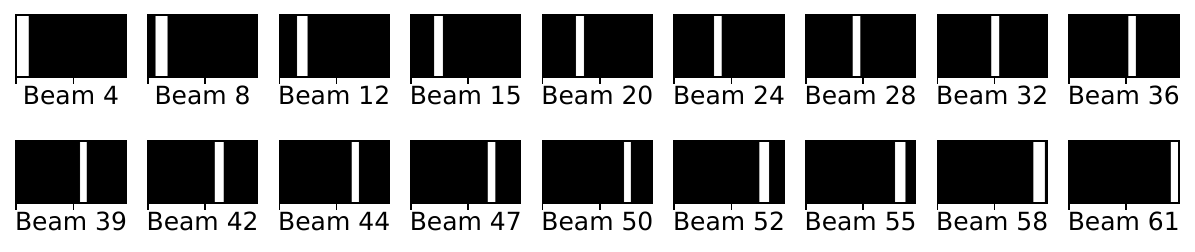}
    \caption{Beams given by the codebook of DeepSense6G dataset \cite{deepsense6g} after mapped on to the image plane.}
    \label{fig:beams_straight}
\end{figure}

This channel encodes beamformed received signal strength measurements directly into the image representation. 
When the camera and BS are co-located with approximately identical horizontal fields of view, each beam's coverage area can be directly projected onto the image plane. The angular sector $[\theta^{(q)}_{\text{start}}, \theta^{(q)}_{\text{end}}]$ of beam $\textbf{f}_q$ as depicted in Fig. \ref{fig:beam_pattern} corresponds to a specific rectangular pixel region in the image whose width is denoted by $[x^{(q)}_{\text{start}}, x^{(q)}_{\text{end}}]$. The transformation between pixel and beam coordinates can be expressed as,
\begin{equation}
\label{eqn:x_region}
    [x^{(q)}_{\text{start}}, x^{(q)}_{\text{end}}] = \frac{W}{\Phi_{\text{end}} - \Phi_{\text{start}}}[(\theta^{(q)}_{\text{start}} - \Phi_{\text{start}}), (\theta^{(q)}_{\text{end}} - \Phi_{\text{start}})],
\end{equation}
where $W$ denotes the image width in pixels. This mapping ensures that each beam direction in the codebook possesses a unique spatial footprint in the image as depicted in Fig. \ref{fig:beams_straight}, forming a set of contiguous or slightly overlapping vertical strips that collectively span the entire image width, 
such that:
\begin{equation}
    \bigcup_{q=1}^Q [x^{(q)}_{\text{start}}, x^{(q)}_{\text{end}}] = [0, W].
\end{equation}

While this mapping establishes the theoretical correspondence between beam directions and image coordinates, practical deployment considerations introduce additional geometric complexities. To circumvent such complexities we design a perspective-aware method that ensures proper spatial alignment between image features and beam directions by accounting for camera geometry. 
The camera used to capture images in the development dataset \cite{deepsense6g} is positioned above street level, where the TX traverses, with a slight downward tilt to ensure comprehensive street coverage. This camera orientation introduces perspective distortion in which vertical structures in the physical environment do not appear vertical in the captured images. 
Instead, these structures exhibit an angular deviation that converges toward a point in the image plane, leading to a fundamental consequence of projective geometry that directly impacts the spatial relationship between visual features and their corresponding beam directions.
To address this geometric distortion, we analyzed the dataset imagery to compute the vertical vanishing point $(x_v, y_v)$,
which is the convergence point in the image plane toward which all physically vertical parallel lines appear to recede due to the camera's perspective projection. Building on this vanishing point analysis, we redesigned the beam mapping strategy described earlier to align with the corrected geometric relationships in the visual data. 

\begin{figure}[!t]
    \centering
    \includegraphics[width=0.48\textwidth]{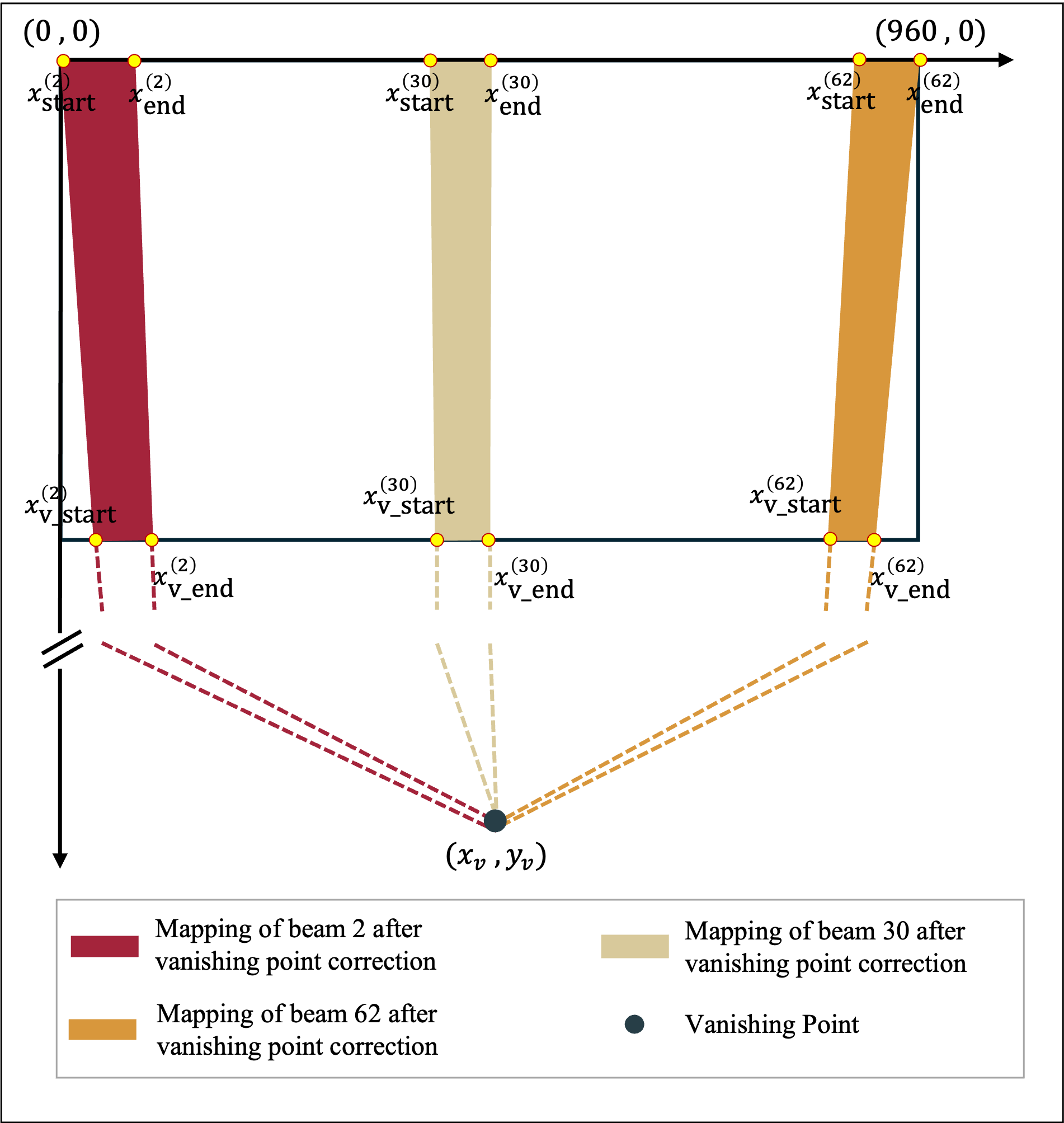}
    \caption{The process of mapping beams given by the codebook of DeepSense6G dataset \cite{deepsense6g} on to the image plane after computing the vanishing point.}
    \label{fig:vp_beams}
\end{figure}

\begin{figure}[!t]
    \centering
    \includegraphics[width=0.48\textwidth]{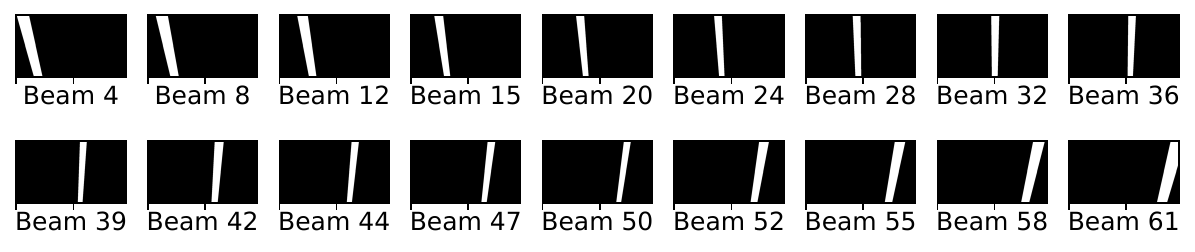}
    \caption{Beams given by the codebook of DeepSense6G dataset \cite{deepsense6g} after mapped on to the image plane based on the vanishing point design.}
    \label{fig:beams_vanishing}
\end{figure}

Once the vanishing point is determined, all vertical lines within the scene should converge through this point. Accordingly, for each beam, the two vertical boundary lines of the beam as illustrated in Fig. \ref{fig:vp_beams} can be derived as,
\begin{equation}
\label{eqn:vp_line}
    y=\frac{y_v}{x_v - x^{(q)}}x - \frac{y_vx^{(q)}}{x_v - x^{(q)}}
\end{equation}
where $(x_v, y_v)$ is the vanishing point in pixel coordinates and \( x^{(q)} \in \{ x^{(q)}_{\text{start}},\; x^{(q)}_{\text{end}} \} \), for any beam $\textbf{f}_q$ resulting in two corresponding lines.
Based on \eqref{eqn:vp_line}, the four corner points defining the new trapezoidal region of projection of the beam on the image for a given beam $q$ can be represented as,
\begin{equation}
\begin{aligned}
&\Big[
\begin{array}{l}
(x^{(q)}_{\text{start}}, 0),\; (x^{(q)}_{\text{end}}, 0), (x^{(q)}_{\text{v\_end}}, W),\; (x^{(q)}_{\text{v\_start}}, W)\;
\end{array}
\Big], \\[6pt]
&x^{(q)}_{\text{v\_end}} = \left(\frac{x_v - x^{(q)}_{\text{end}}}{y_v} H + x^{(q)}_{\text{end}},\; W\right), \\[6pt]
&x^{(q)}_{\text{v\_start}} = \left(\frac{x_v - x^{(q)}_{\text{start}}}{y_v} H + x^{(q)}_{\text{start}},\; W\right),
\end{aligned}
\end{equation}
where $H, W$ is the height and width of the image in pixel coordinates. By using the vanishing point as a geometric reference, we derive a rectified beam-to-pixel mapping as illustrated in Fig. \ref{fig:beams_vanishing} that compensates for the projective distortion.

\subsubsection{Structuring Channel 3: Zero-Padded Channel for Transfer Learning Compatibility}
The third channel of our input representation consists entirely of zero-valued pixels. Although it does not directly contribute to the detection task, its inclusion serves a critical architectural purpose, maintaining dimensional compatibility with pretrained models.
The YOLOv10 architecture is pretrained on the MS COCO dataset \cite{cocodataset}, which uses standard RGB images. To leverage these pretrained weights, our input must conform to this three-channel format. By constructing our input with the mmWave power profile channel, the visually-ablated scene channel, and a zero-padded third channel, we preserve compatibility with the pretrained network while introducing domain-specific information.
This transfer learning approach is essential given the limited scale of the DeepSense6G dataset \cite{deepsense6g} as training a deep detection network from scratch would require substantially more labeled samples than available.
The zero-valued third channel enables effective transfer learning without impacting the model's ability to learn mmWave-visual correlations during fine-tuning.

Once all three channels are compiled, we feed this new image into the object detector.
The primary task of the detector is to learn an optimal mapping function $g$ that establishes the correspondence between multimodal sensory inputs and TX location:
\begin{equation}
    g: \{[I(t),P(t)] \rightarrow bbox_{TX\_p}(t)\},
\end{equation}
where $I(t)$ denotes the image captured and $P(t)$ represents the beamformed power measurements across all the beams of the codebook at time $t$. The output $bbox_{TX\_p}(t)$ constitutes the predicted bounding box (bbox) encapsulating the TX location at the corresponding time instant. This bbox is parameterized by four scalar values with $x$ and $y$ representing the pixel coordinates of the box centroid in the image plane, and $w$ and $h$ defining the width and height of the bbox, respectively. To localize the TX bbox, we utilize the YOLOv10 object detector \cite{yolov10}, which processes our resulting three-channel composite input.

To evaluate the TX identification performance, we utilize Scenarios 3 and 4 from the DeepSense 6G dataset \cite{deepsense6g}. To train the above model we split the Scenario 3 data into a 70:30 train-test split. We consider these scenarios as they represent the most challenging operational conditions in the dataset, characterized by nighttime environments with rapidly moving TXs, where the low-light imaging conditions introduce significant visibility constraints that rigorously evaluate our proposed method. The trained model is evaluated on the held-out portion of Scenario 3 and the complete Scenario 4.
The training set comprises 1204 samples, while testing is performed on 288 samples from Scenario 3, and 275 samples from Scenario 4.
Detection accuracy is calculated using the Intersection over Union (IoU) metric between predicted and ground truth bboxes as follows:
\begin{equation}
    \mathsf{Acc.} = \frac{\sum_D\mathbb{I}\{\mathsf{IoU}(bbox_{TX\_p}(t), bbox_{TX\_gt}(t)) \geq Z\}}{D},
\end{equation}
where $\mathbb{I}$ denotes the indicator function, $D$ represents the total number of test samples, and $\mathsf{IoU}(bbox_{TX\_p}(t), bbox_{TX\_gt}(t))$ measures the spatial overlap between the predicted bbox $bbox_{TX\_p}(t)$ and the ground truth annotation $bbox_{TX\_gt}(t)$ at time $t$. The threshold $Z$ determines the minimum IoU value required for a detection to be considered correct. A prediction is deemed accurate when the overlap between predicted and ground truth boxes exceeds this threshold, indicating successful TX localization. 

\begin{table}[t]
\caption{DeepSense 6G TX Identification Accuracy: Models trained on Scenario 10 and tested on Scenarios 10, 11 and 12.}
\centering
\resizebox{0.5\textwidth}{!}{%
\begin{tabular}{p{0.02cm}p{5cm}p{0.75cm}p{0.75cm}p{0.7cm}}
\toprule
 & \multirow{2}{*}{\textbf{Methodology}} & \multicolumn{3}{l}{\textbf{TX Identification Accuracy}} \\
    &   & $M$=1   & $M$=3  & $M$=5 \\
\midrule
\multirow{5}{*}{\rotatebox{90}{Scenario 3}}& Charan \textit{et al.} \cite{charan23} & 98.43\% & 99.00\% & 99.48\% \\
& \textbf{Proposed: TX Identification Method 1} & \textbf{99.65\%} & \textbf{100.00\%} & \textbf{100.00\%}\\
& \textbf{Proposed: TX Identification Method 2} & \textbf{96.52\%} & \textbf{100.00\%} & \textbf{100.00\%}\\
& Ablation study 1: RGB input & 99.10\% & 100.00\% & 100.00\% \\
& Ablation study 2: All zero mmWave channel & 24.92\% & 25.61\% & 26.82\%\\
\midrule
\multirow{5}{*}{\rotatebox{90}{Scenario 4}}& Charan \textit{et al.} \cite{charan23} & 98.43\% & 99.00\% & 99.48\% \\
& \textbf{Proposed: TX Identification Method 1} & \textbf{99.65\%} & \textbf{100.00\%} & \textbf{100.00\%}\\
& \textbf{Proposed: TX Identification Method 2} & \textbf{96.52\%} & \textbf{100.00\%} & \textbf{100.00\%}\\
& Ablation study 1: RGB input & 99.10\% & 100.00\% & 100.00\% \\
& Ablation study 2: All zero mmWave channel & 24.92\% & 25.61\% & 26.82\%\\
\bottomrule
\end{tabular}%
\label{tab:tx_identi_results}
}
\end{table}

In Scenarios 3 and 4 of the DeepSense 6G dataset, the TX across all video frames is consistently the same vehicle equipped with the TX. Without the preprocessing steps that remove visual information, an object detector would inadvertently learn car-specific characteristics such as shape and color patterns to identify the TX, rather than relying on the mmWave channel measurements. This would result in a model that bases its detection on visual appearances rather than the underlying mmWave propagation characteristics, essentially creating a scenario-specific solution that fails to generalize to different objects. 

Table \ref{tab:tx_identi_results} presents a comprehensive performance comparison between our proposed approach, demonstrating the effectiveness of visual ablation and the superior detection accuracy achieved through mmWave-driven identification. All results are evaluated using an IoU threshold of $Z=0.5$.
To enhance detection robustness, our TX identification framework uses information from $M$ consecutive frames, as illustrated in Fig. \ref{fig:framework1}. This approach reduces the impact of occasional missed detections or false positives in individual frames by checking for consistency across multiple frames.
The object most frequently classified as the TX across these $M$ frames is designated as the final detection. We evaluate this approach with $M \in \{1, 3, 5\}$ frames.
To validate that our model genuinely relies on mmWave power profiles rather than residual visual features, we conduct two ablation studies as follows: 
\begin{enumerate}
    \item A baseline YOLOv10 model trained on unmodified RGB images with full color and shape information.
    \item A control configuration where the mmWave power profile channel is replaced with zeros, effectively removing all channel information.
\end{enumerate}
The results, as presented in Table \ref{tab:tx_identi_results}, reveal that without mmWave channel input, detection accuracy degrades substantially, confirming that visual information alone is insufficient for reliable TX identification in our framework. Conversely, the marginal difference in accuracy between models trained on original RGB images versus our visually-ablated inputs demonstrates that shape and color contribute negligibly to detection performance. This validates our central hypothesis, which is that accurate TX identification is driven primarily by the encoded mmWave power profile rather than appearance-based features, ensuring generalization across different TX objects.

\subsection{Transmitter Tracking}
\label{sec:TX_track}

Next, we transition to continuous spatial tracking of the TX within the beam steering range of the BS. The main purpose of the tracker is to maintain precise localization of the detected TX across temporal sequences. The main objective of incorporating an object-tracking mechanism is to eliminate the need for repeated EBS at each time instance. Traditional beam management requires periodic EBS operations to locate and serve the TX, which introduces substantial latency and computational burden, which is the primary challenge that vision-aided beamforming seeks to address. Since our framework reliably identifies the TX, continuous tracking enables direct beam steering toward the tracked TX without resorting to exhaustive scanning or other computationally intensive procedures in subsequent frames. The tracker provides real-time location of the TX, allowing the system to adjust beam direction based on the said TX location. This beam prediction methodology based on image analysis is detailed in Section \ref{sec:beam_pred}, which fundamentally transforms beam management from a computationally intensive search problem into an efficient vision-guided targeting operation.

For the tracking phase, we employ the DeepOCSort \cite{deepOCSort} object tracker to maintain continuous TX localization across consecutive frames. Similar to the TX identification stage, we deploy an object detection model operating on standard RGB images to detect all objects within the scene, irrespective of their classification as TX or distractors. Upon successful TX identification in the initial frame, we initialize the tracker with the bbox coordinates of the detected TX. The tracker then receives updated bbox detections from subsequent frames, enabling it to maintain the TX trajectory throughout its presence in the coverage area. Notably, unlike the TX identification phase, where shape and color masking were employed, we retain all visual information in this tracking stage. 
The tracker leverages these rich visual features to maintain accurate TX association across frames, ensuring reliable spatial localization for downstream beam prediction operations.

\subsection{Beam Prediction}
\label{sec:beam_pred}
This subsection presents a novel dual-stage methodology for determining the top-$N$ candidate beams serving the detected and monitored TX, as elaborated in Sections \ref{sec:TX_identi} and \ref{sec:TX_track}. This beam prediction framework constitutes the final component of our vision-aided beamforming architecture.

\subsubsection{Beam Search Space Reduction}
\label{sec:beam_search_space_red}
The first stage focuses on constraining the beam search space through geometric analysis and visual isolation of the TX. 
First, we perform spatial isolation of the TX within the captured image frame by eliminating all external and distractor objects and background elements. This isolation procedure utilizes the object mask of the TX provided by the tracking system from Section \ref{sec:TX_track}, ensuring that only pixels corresponding to the TX are retained in the processed image.
Following TX isolation, leveraging the vertical vanishing point calibration and redesigned beam shape projections described in Section \ref{sec:TX_identi}, we systematically overlay each perspective-corrected beam shape from the redesigned codebook onto the isolated TX region. For each beam projection, we perform pixel-level intersection analysis to determine spatial overlap with the TX pixels. Beams demonstrating at least one overlapping pixel with the isolated TX region are retained, forming a reduced candidate beam subset $\mathcal{B}_{\text{reduced}} \subset \mathcal{C}_r$.

\subsubsection{Top-N Beam Prediction}
\begin{figure}[t]
    \centering
    \includegraphics[width=0.5\textwidth]{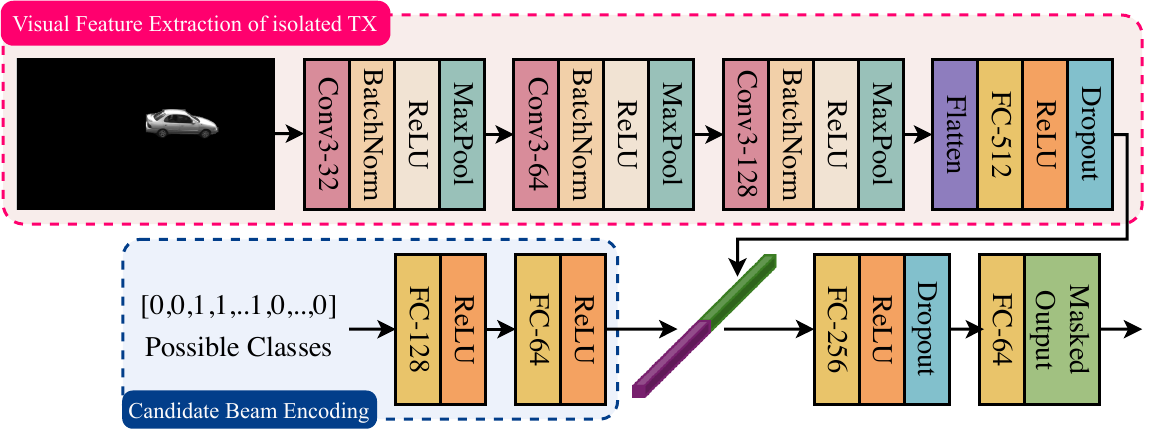}
    
    \caption{Proposed dual-branch neural network architecture for top-$N$ beam selection, integrating Visual Feature Extraction of isolated TX (pink) with Candidate Beam Encoding (blue).}
    \label{fig:nn}
\end{figure}

We develop a specialized neural network architecture to identify the top-$N$ optimal beam indices serving the tracked TX. The design of this network, as depicted in Fig. \ref{fig:nn}, employs a dual-branch architecture:
\begin{itemize}
    \item \textbf{Visual Feature Extraction: } This branch accepts the isolated TX image as input and processes it through three successive convolutional blocks. Each block comprises a Convolutional layer (Conv), Batch Normalization layer (BatchNorm), and Rectified Linear Unit (ReLU) activation function, succeeded by a Max Pooling layer (MaxPool). The resulting feature maps are subsequently flattened into a one-dimensional representation, enabling fusion with the auxiliary branch output.
    \item \textbf{Candidate Beam Encoding: } This branch handles the constrained beam search space obtained from the Section \ref{sec:beam_search_space_red}, where candidate beam indices are represented using binary encoding (``1'' for viable beams, ``0'' otherwise). This encoded vector traverses two successive fully connected (FC) layers for feature extraction. The output features from this branch are merged with the flattened visual features through concatenation.
\end{itemize}
Following the fusion of both feature streams, the consolidated representation undergoes additional refinement through two FC layers. The processed features then feed into a 64-class classification head, corresponding to the complete beam codebook size in the dataset, that determines the top-$N$ beam selections. To enforce geometric consistency with the reduced search space, we apply a post-classification masking operation that suppresses predictions outside the geometrically-feasible beam subset by zeroing the corresponding output weights.

\begin{table}[t]
\caption{Beam Prediction Accuracy: Models trained on Scenario 3 and tested on both Scenarios 3 and 4.}
\centering
\resizebox{0.5\textwidth}{!}{%
\begin{tabular}{p{0.02cm}p{5cm}p{0.75cm}p{0.75cm}p{0.7cm}}
\toprule
 & \multirow{2}{*}{\textbf{Methodology}} & \multicolumn{3}{l}{\textbf{Beam Prediction Accuracy}} \\
    &   & Top-$1$   & Top-$3$  & Top-$5$ \\
\midrule
\multirow{4}{*}{\rotatebox{90}{Scenario 3}}& Proposed & \textbf{59.72\%} & \textbf{89.24\%} & \textbf{98.96\%} \\
& Ablation study 1: &  7.29\% & 16.67\% & 22.92\%\\
& Ablation study 2: & 56.94\%  & 87.67\% & 98.26\% \\
& Ablation study 3: & 51.39\% & 81.94\% & 93.40\% \\
\midrule
\multirow{4}{*}{\rotatebox{90}{Scenario 4}}& Proposed & \textbf{57.83\%} & \textbf{84.42\%} & \textbf{96.01\%}\\
& Ablation study 1: & 6.82\% & 14.34\% & 21.98\% \\
& Ablation study 2: & 55.89\% & 82.23\% & 95.74\% \\
& Ablation study 3: & 50.45\% & 78.28\% & 91.77\%\\
\bottomrule
\end{tabular}%
\label{tab:beam_pred_results}
}
\end{table}

Table \ref{tab:beam_pred_results} presents the performance evaluation of our proposed beam prediction methodology. To demonstrate the contribution of each architectural component, we perform three systematic ablation experiments:
\begin{itemize}
    \item Ablation Experiment 1: This experiment examines the effect of TX isolation by substituting the isolated TX region with the complete input image containing all environmental objects and background clutter. This variant assesses whether selective TX focus improves beam prediction accuracy.
    \item Ablation Experiment 2: This configuration eliminates the candidate beam encoding branch, along with the corresponding binary search space vector and post-classification masking layer. This experiment quantifies the benefit of geometric search space constraints on prediction performance.
    \item Ablation Experiment 3: This variant replaces the perspective-corrected beam projections with the original uncorrected beam shapes, thereby excluding vanishing point compensation. This experiment evaluates the necessity of geometric beam shape transformation for effective beam search space reduction.
\end{itemize}
The ablation study outcomes, detailed in Table \ref{tab:beam_pred_results}, demonstrate that each methodological component provides substantial performance gains. These findings validate our integrated design approach, emphasizing that the combination of TX-focused image processing, geometrically-constrained search space, and perspective-aware beam projection yields superior beam prediction accuracy through complementary mechanisms.
Network training used the cross-entropy loss function with an initial learning rate of 0.001 and the Adam optimizer across 30 training epochs. We applied a learning rate decay of 0.0001 upon achieving 59\% validation accuracy.

\subsection{Numerical Results}
\label{sec:num_res}

\begin{figure}[t!]
    \centering
    \subfloat[Top-N Beam Prediction criteria for Scenario 3]{
        \includegraphics[width=0.48\textwidth]{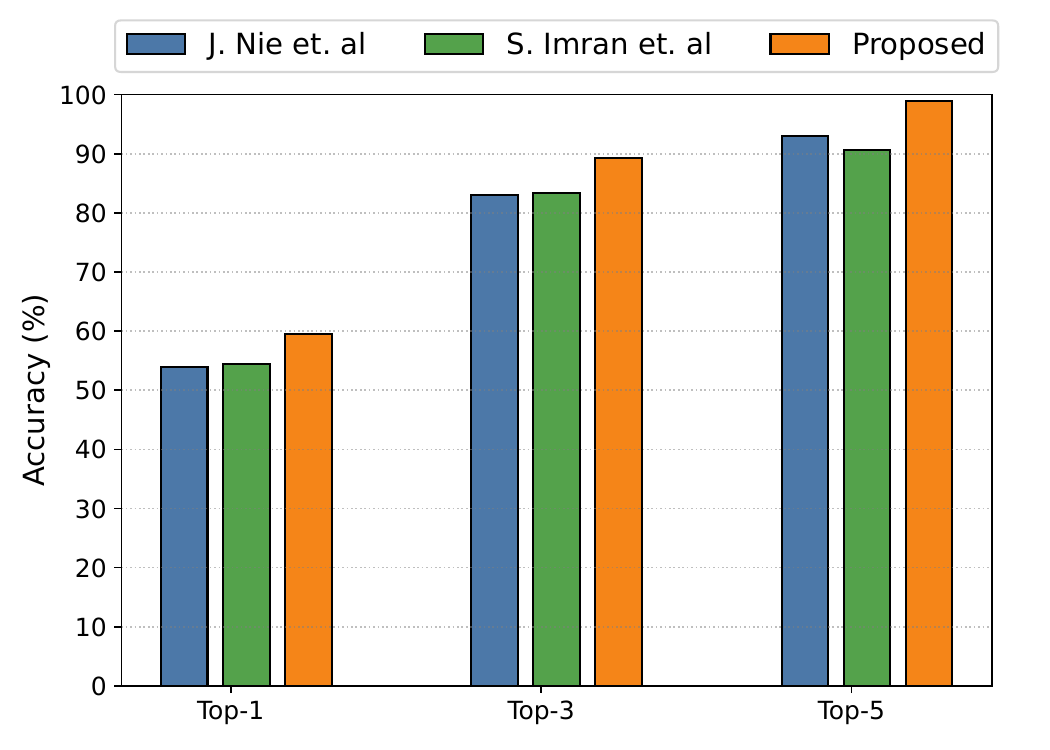}
        \label{fig:results-scne3}
    }\hfill
    \subfloat[Top-N Beam Prediction criteria for Scenario 4]{
        \includegraphics[width=0.48\textwidth]{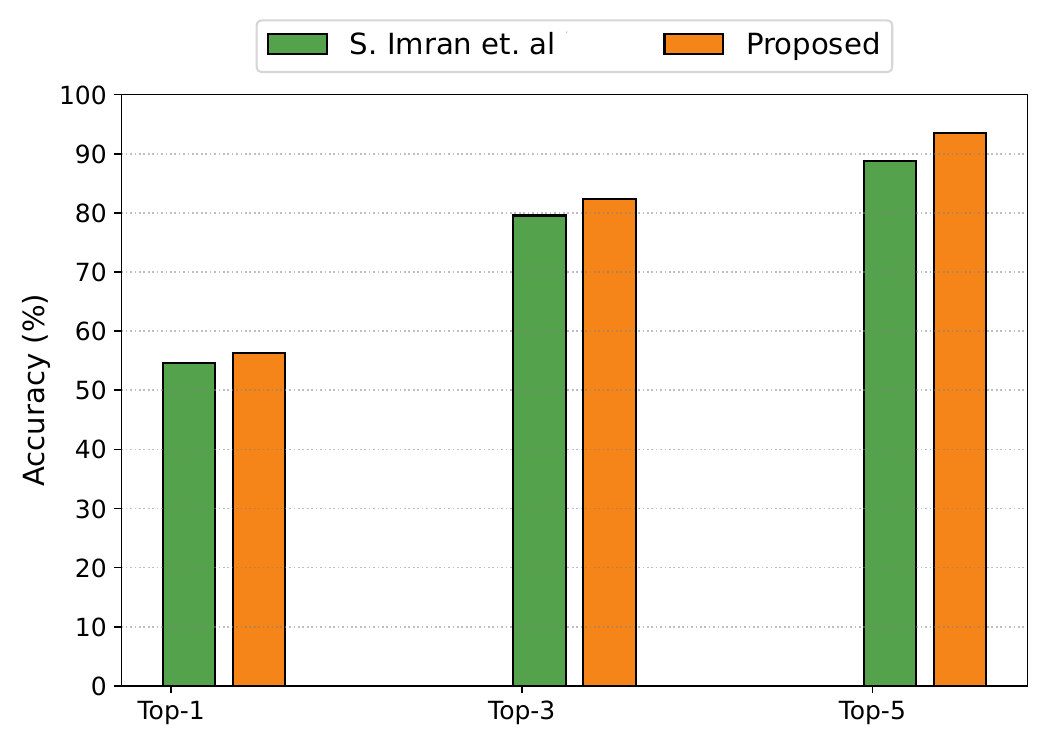}
        \label{fig:results-scen4}
    }
    \vspace{1em}
    \caption{Comparison of Top-N Beam Prediction metrics with current state-of-the-art methods. ``J. Nie et. al'' represents the work done in \cite{nie2023} and ``S. Imran et. al'' represents the results generated by applying the method in \cite{imran23}.}
    \label{fig:results}
\end{figure}

In the preceding Sections \ref{sec:TX_identi} and \ref{sec:beam_pred}, we analyzed the performance of individual pipeline components. This section evaluates the complete end-to-end system by integrating all three stages and benchmarking the overall performance against current state-of-the-art methodologies. To ensure fair comparison, our proposed framework and all baseline methods were evaluated on identical hardware configurations, thereby attributing observed performance improvements solely to algorithmic innovations rather than computational advantages. We evaluate this framework on Scenarios 3 and 4 of the DeepSense 6G dataset \cite{deepsense6g}. These nighttime scenarios impose severe visual degradation and sensor noise, and our method establishes new performance benchmarks under these challenging conditions.

Fig. \ref{fig:results} presents a comprehensive performance comparison between our framework and the baseline methods, which are the approach from \cite{imran23} and the hybrid vision-position technique from \cite{nie2023} that leverages supplementary GPS positioning data. For Scenario 3, illustrated in Fig. \ref{fig:results-scne3}, our method delivers accuracies of 59.51\%, 88.93\%, and 98.61\% for the Top-1, Top-3, and Top-5 prediction metrics, respectively, as specified in Section \ref{sec:sysModel}. Similarly, the results of scenario 4 shown in Fig. \ref{fig:results-scen4} demonstrate accuracies of 56.36\%, 82.27\%, and 93.57\% for the corresponding metrics.
Our complete pipeline demonstrates consistent performance gains of at least 6\% over competing methods across all evaluated beam prediction metrics.
Remarkably, our purely vision-based approach achieves performance levels comparable to the multi-modal GPS-augmented method from \cite{nie2023}, validating the robustness of our technique despite operating without auxiliary positioning sensors.

\section{Vision-aided Blockage Identification}
\label{sec:va-bi}

\begin{figure*}[t]
\vspace*{0.1in}
    \centering
    \includegraphics[width=1\textwidth]{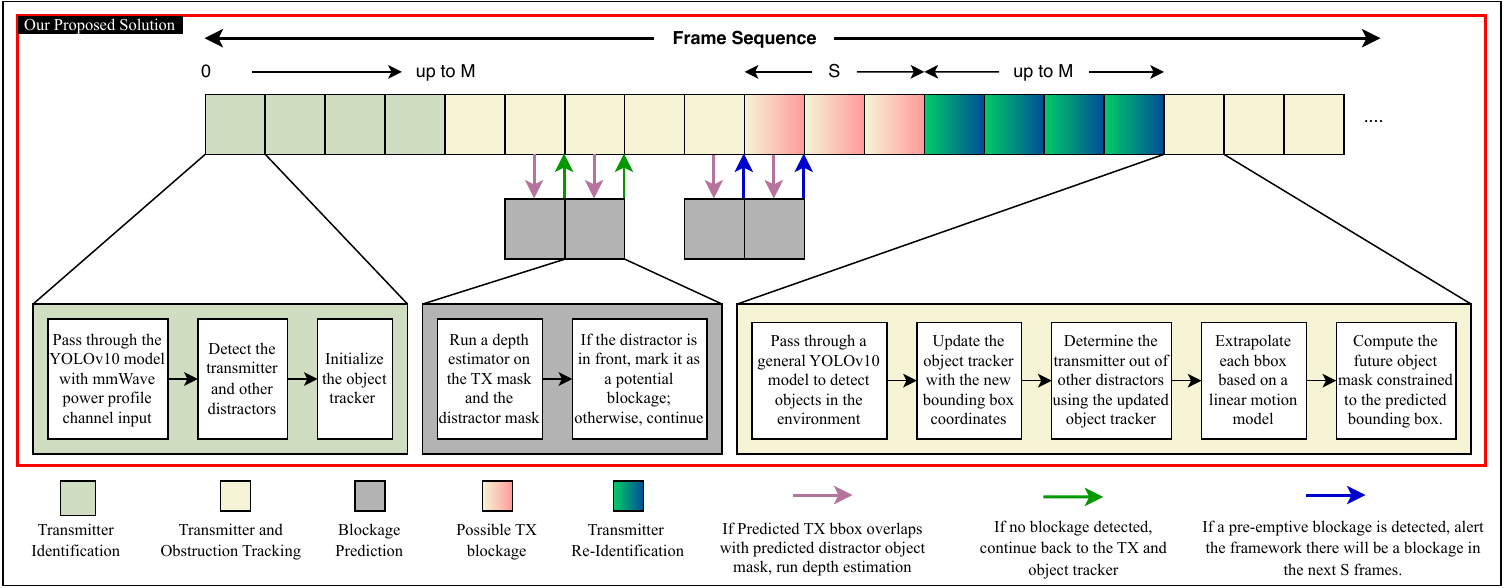}
    \caption{Proposed end-to-end vision-aided blockage prediction framework (red box). The first $M$ frames identify the TX, followed by tracking the TX and all objects in subsequent frames. Bboxes are extrapolated to predict future positions, and potential blockages are detected when predicted distractor object masks overlap with the predicted TX box. Upon overlap detection, a depth estimator determines if the distractor is positioned in front of the TX. If confirmed, a blockage is predicted within the next $S$ frames. After $S$ frames, TX re-identification localizes the TX post-blockage, and the process repeats.}
    \label{fig:framework2}
\end{figure*}

This section presents our novel three-stage framework that leverages RGB imagery and mmWave power profiles for accurate LoS blockage prediction in dynamic mmWave communication environments. Our framework is the first systematic framework to predict LoS blockages in mmWave systems with continuously mobile TXs, addressing a critical gap in existing literature that exclusively considers static TX deployments. The framework,  as illustrated in Fig. \ref{fig:framework2}, integrates mmWave and visual data through a sequential processing pipeline designed to achieve pre-emptive blockage prediction capabilities in highly dynamic operational environments. 
The remainder of this section proceeds by detailing each component of the framework as follows, followed by an overall performance evaluation.
\begin{enumerate}
    \item \textbf{TX Detection (Section \ref{sec:TX_identi_bl})}: This stage follows the methodology described in Section \ref{sec:TX_identi}, focusing on accurate TX identification and localization within the coverage area.
    \item \textbf{TX and Obstruction Tracking (Section \ref{sec:TX_track_bl})}: Temporal tracking of the detected TX and potential obstructing entities simultaneously, while predicting future trajectories by employing state-space estimation techniques. 
    \item \textbf{Blockage Prediction and Fast Recovery (Section \ref{sec:blockage})}: Pre-emptive blockage prediction by depth estimation and rapid beam recovery for proactive beam management.
\end{enumerate}
A critical distinction of our proposed blockage prediction framework is that it operates entirely without training data beyond the initial TX identification neural network, where all blockage prediction, tracking, depth estimation, and recovery mechanisms are evaluated on completely unseen data that the framework has never encountered during any training phase.

\subsection{Transmitter Identification}
\label{sec:TX_identi_bl}

The initial stage of this framework mirrors the TX identification methodology presented in Section \ref{sec:TX_identi}, serving as a critical prerequisite for blockage prediction. Accurate detection and localization of TX are essential for assessing potential LoS obstructions that may occur between the TX and the BS during mobile operation. For the blockage prediction framework, we employ Scenarios 10, 11, and 12 from the DeepSense 6G dataset \cite{deepsense6g}. These scenarios are specifically selected due to their high prevalence of dynamic blockage events involving moving TXs, which is a challenging operational regime that remains largely unexplored in existing literature. 
Unlike prior studies that primarily address static blockage conditions or stationary TX configurations, these scenarios capture frequent obstruction events as TXs traverse the coverage area, presenting realistic mobility-induced blockage patterns. 
To the best of our knowledge, this work represents the first systematic investigation of TX identification for these scenarios, thereby establishing initial performance benchmarks for this dataset.

We apply the same visual feature suppression strategy detailed in Fig. \ref{fig:tx_identi} to enforce reliance on mmWave power profile information rather than visual appearance features. For model training, we partition Scenario 10 using a 70:30 train-test split. We then evaluate the trained model on the remaining Scenario 10 test partition and on the complete Scenarios 11 and 12, yielding a total of 348 training samples and 793 test samples across all scenarios.

\begin{table}[t]
\caption{DeepSense 6G TX Identification Accuracy: Models trained on Scenario 10 and tested on Scenarios 10, 11 and 12.}
\centering
\resizebox{0.5\textwidth}{!}{%
\begin{tabular}{p{0.02cm}p{5cm}p{0.75cm}p{0.75cm}p{0.7cm}}
\toprule
 & \multirow{2}{*}{\textbf{Methodology}} & \multicolumn{3}{l}{\textbf{TX Identification Accuracy}} \\
    &   & $M$=1   & $M$=3  & $M$=5 \\
\midrule
\addlinespace[1ex]
\multirow{4}{*}{\rotatebox{90}{Scenario 10}}%
& \textbf{Ours: TX Identification Method 1} & \textbf{97.59\%} & \textbf{100.00\%} & \textbf{100.00\%}\\
& \textbf{Ours: TX Identification Method 2} & \textbf{98.80\%} & \textbf{100.00\%} & \textbf{100.00\%}\\
& Ablation study 1: RGB input & 98.80\% & 100.00\% & 100.00\% \\
& Ablation study 2: All zero mmWave channel & 24.92\% & 25.61\% & 26.82\%\\

\addlinespace[1ex]
\midrule
\addlinespace[1ex]
\multirow{4}{*}{\rotatebox{90}{Scenario 11}}& \textbf{Ours: TX Identification Method 1} & \textbf{94.03\%} & \textbf{99.37\%} & \textbf{99.68\%}\\
& \textbf{Ours: TX Identification Method 2} & \textbf{94.03\%} & \textbf{99.37\%} & \textbf{99.68\%}\\
& Ablation study 1: RGB input & 95.28\% & 99.68\% & 99.68\% \\
& Ablation study 2: All zero mmWave channel & 22.32\% & 23.80\% & 24.95\%\\

\addlinespace[1ex]
\midrule
\addlinespace[1ex]
\multirow{4}{*}{\rotatebox{90}{Scenario 12}}
& \textbf{Ours: TX Identification Method 1} & \textbf{95.47\%} & \textbf{99.62\%} & \textbf{99.62\%}\\
& \textbf{Ours: TX Identification Method 2} & \textbf{95.47\%} & \textbf{99.62\%}& \textbf{99.62\%}\\
& Ablation study 1: RGB input & 98.11\% & 100.00\% & 100.00\%\\
& Ablation study 2: All zero mmWave channel & 22.89\% & 23.72\% & 24.45\%\\
\addlinespace[1ex]
\bottomrule
\end{tabular}%
\label{tab:tx_identi_results_bl}
}
\end{table}

Table \ref{tab:tx_identi_results_bl} presents the TX identification performance across Scenarios 10, 11, and 12, demonstrating the robustness of our proposed framework despite the fundamental change in TX carrier type from vehicular mounting (Scenarios 3 and 4) to pedestrian handheld operation. Both proposed methods consistently achieve accuracies exceeding 94\% when utilizing a single observation frame ($M=1$) and approach near-perfect accuracy when leveraging three or five consecutive frames ($M=3$ or $M=5$) across all evaluation scenarios, effectively detecting the TX amidst numerous similar moving objects under diverse environmental conditions. The results of the ablation study reinforce the observations from Section \ref{sec:TX_identi}, namely that while RGB-only input maintains competitive performance, the complete removal of mmWave channel information results in substantial accuracy degradation. 
This performance pattern re-confirms that the model predominantly relies on mmWave power profile characteristics for TX identification. 

\subsection{Transmitter and Obstruction Tracking}
\label{sec:TX_track_bl}

Following successful TX identification, the framework transitions to continuous spatial tracking of both the TX and surrounding objects within the BS's coverage area. This tracking stage serves a dual purpose: maintaining precise localization of the TX across temporal sequences while simultaneously predicting potential obstructions that may interrupt the LoS communication link. In dynamic urban environments, mobile objects such as pedestrians, vehicles, and other moving entities can traverse the propagation path between the BS and the TX, causing intermittent blockages that severely degrade mmWave link quality. 
To address this challenge, we propose a custom tracking pipeline built upon the Deep OC-SORT framework \cite{deepOCSort}, which we enhance with domain-specific modifications tailored to the mmWave communication context. 
This subsection details the tracking methodology, trajectory management, and the path intersection that collectively lead to proactive blockage detection in dynamic environments.

Our proposed tracker assigns a unique identification number to each detected object from the TX identification stage, enabling consistent identity maintenance across temporal sequences. Having successfully identified the TX, we leverage this unique ID to continuously track its trajectory within the BS's coverage area without requiring exhaustive beam sweeping at each time instance to locate the TX and proactively predict the most optimal beam as detailed in Section \ref{sec:beam_pred}. 
While trajectory tracking can be used to efficiently steer the beam toward the TX, it does not address dynamic LoS blockages caused by moving objects in the environment, a critical challenge in mmWave communications. Through our proposed tracker, we aim to anticipate future LoS blockages between the BS and the tracked TX, providing sufficient time to execute proactive mitigation strategies. 

Our proposed tracking algorithm maintains a history of bbox coordinates for each tracked object across consecutive frames. For each object $i$ at time $t$, we record the complete bbox state $bbox_i(t) = [x_1^i(t), y_1^i(t), x_2^i(t), y_2^i(t)]$, where $(x_1, y_1)$ and $(x_2, y_2)$ denote the top-left and bottom-right corners of the bbox, respectively. 
To forecast future object positions, we employ a linear motion model based on the aggregation of the most recent trajectory history comprising the most recent $F$ frames. 
Our prediction operates on discrete frame indices rather than continuous-time coordinates. The bbox predicted for frame $f+s$, where $s$ represents the number of frames into the future from the current frame $f$, is computed as
\begin{equation}
\label{eqn:pred_bbox}
    bbox_i(f+s) = bbox_i(f) + s \cdot \Delta{bbox}_i(f),
\end{equation}
where $\Delta{bbox}_i(t)$ represents the average displacement vector computed from recent history as
\begin{equation}
    \Delta{bbox}_i(f) = \frac{1}{F-1} \sum_{k=1}^{F-1} [bbox_i(f-k+1) - bbox_i(f-k)],
\end{equation}
Having predicted the future bbox coordinates, we generate corresponding object masks for each object inside the predicted bbox. The current segmentation mask of object $i$ at frame $f$, denoted by $M_i(f)$, is transformed and fitted into the predicted bbox, $bbox_i(f+s)$, through spatial scaling and translation operations. This yields the predicted mask $\hat{M_i}(f+s)$, which preserves the object's shape characteristics while repositioning it according to the forecasted trajectory, which will be essential for the next stage of blockage prediction. 

Scenarios 10, 11, and 12 of the DeepSense6G dataset \cite{deepsense6g} exhibit frequent frame drops and irregular frame intervals, which pose significant challenges for maintaining consistent object identities across sequences. When frames are missing, standard tracking algorithms often fail to re-associate detected objects with their previous identities, instead assigning new tracker IDs to the same physical objects. This ID fragmentation is particularly problematic for our framework, as it would necessitate repeated TX identification, which is the computationally expensive process we seek to avoid. To address this limitation, we develop an ID reassociation mechanism that extends the Deep OC-SORT tracker with gap-bridging capabilities. 
When frame discontinuity is detected, indicated by abrupt changes in assigned tracker IDs between consecutive processed frames, we instantiate a parallel candidate tracker that monitors objects with newly assigned IDs while maintaining the original persistent tracker that contains historical trajectories with established IDs. For each object $i$ in the persistent tracker, we leverage the motion model from \eqref{eqn:pred_bbox} to predict an intermediate bbox position corresponding to the estimated location during the missing frame interval. We then compute the IoU between each interpolated bbox from the persistent tracker and all current detections with new IDs in the candidate tracker. Using a greedy matching strategy, we sequentially assign new IDs to old IDs based on maximum IoU overlap, ensuring one-to-one correspondence. Objects that achieve IoU above a threshold $\tau_{\text{match}}$ are considered successful re-associations. Once ID correspondences are established, we update the persistent tracker by appending both the interpolated frame and current frame observations to the historical trajectory and maintain the old tracking IDs relevant to the persistent tracker. The candidate tracker is then discarded, and tracking continues with the consolidated persistent tracker. This mechanism preserves the TX's identity throughout frame gaps, eliminating the need for repeated TX identification. 

Following successful TX and multi object tracking, we analyze the spatial relationship between the predicted TX bbox and the predicted distractor object masks to assess potential blockage scenarios. We employ the TX bbox rather than its segmentation mask due to the physical configuration of the mmWave antenna. The pedestrian carrier holds the TX unit extended away from their body, and relying solely on the person's segmented pixels would fail to encompass the actual antenna location and its spatial extent between the hand and the device. Conversely, for distractor objects, we utilize segmentation masks instead of bboxes to determine potential obstructions between the TX and BS. This asymmetric approach is critical for minimizing false positive blockage predictions, as distractor bboxes inherently encompass significantly larger spatial regions than the actual object occupancy. Using bboxes for distractors would introduce extraneous background and environmental pixels at the box periphery, leading to false overlap detections that would propagate erroneous depth information to subsequent processing stages and compromise the accuracy of our depth estimation module.

We formalize the overlap detection criterion as follows. Let $B_{TX}$ denote the bbox region of the predicted TX, and $M_i$ represent the segmentation mask of the $i$-th distractor object. The spatial overlap $O_i$ between the TX and the $i$-th distractor is computed as:
\begin{equation}
O_i = |B_{TX} \cap M_i|
\end{equation}
where $|\cdot|$ denotes the pixel count of the intersection region. If $O_i > 0$ for any distractor $i$, we flag that particular distractor as a possibility for a potential blockage candidate. For each identified blockage scenario, we extract the corresponding TX bbox region $B_{TX}$ and the overlapping distractor mask $M_i$ for subsequent depth estimation, as detailed in Section \ref{sec:blockage}.

\subsection{Blockage Prediction and Fast Recovery}
\label{sec:blockage}

In this section, we determine whether a predicted overlapping distractor object actually resides in front of the TX, as only such configurations result in LoS blockage. Distractors positioned behind the TX do not obstruct the communication link and thus require no mitigation. To establish the relative depth ordering, we employ the MiDaS \cite{midas} depth estimation model to analyze the current frame containing both the flagged distractor and the TX, computing their respective depth values to ascertain which object is closer to the BS. Critically, depth estimation is invoked only when spatial overlap has been detected in the preceding stage, ensuring computational efficiency by avoiding unnecessary depth calculations.

Rather than computing a simple arithmetic mean of depth values across object regions, we employ a centroid-weighted averaging that accounts for the spatial reliability characteristics of segmentation masks and depth measurements. This approach is motivated by several critical observations regarding depth estimation quality: The boundary regions of human segmentation masks are inherently prone to errors, including artifacts from hair strands, clothing edges, and background pixel leakage due to imperfect segmentation. Depth sensors similarly produce unstable and inaccurate depth estimates at object boundaries, where depth discontinuities and mixed pixels compromise measurement fidelity. A uniform mean across all mask pixels assigns equal importance to both the stable central body regions and these unreliable peripheral measurements, thereby distorting the representative depth value. In contrast, the central torso region exhibits smooth, consistent depth values that accurately represent the primary body mass, while elongated limbs extending toward or away from the camera introduce depth variations that do not reflect the core object position. Furthermore, partial occlusions and segmentation imperfections predominantly affect edge regions, making boundary pixels particularly susceptible to errors.

The weighted mean approach directly addresses these limitations by emphasizing high-quality central depth measurements while suppressing noisy boundary contributions. The weighting procedure operates as follows. First, we compute the spatial centroid $(c_x, c_y)$ of the mask of each TX and distractor object region.
For each pixel within the mask, $(x,y)$, we compute its Euclidean distance $d(x,y)$ from the centroid.
We then apply a Gaussian weighting function, $w(x,y)$, to assign higher weights to pixels near the centroid and progressively lower weights to peripheral pixels,
\begin{equation}
w(x,y) = \exp\left(-\frac{d(x,y)^2}{2\sigma^2}\right)
\end{equation}
where $\sigma$ controls the spatial extent of the weighting function. Finally, the weighted mean depth $\bar{D}$ is computed as:
\begin{equation}
\bar{D} = \frac{\sum_{(x,y) \in M} w(x,y) \cdot D(x,y)} {\sum_{(x,y) \in M} w(x,y)}
\end{equation}
where $D(x,y)$ represents the depth value at pixel $(x,y)$. 
If any or all predicted distractors satisfy $\bar{D}_{\text{distractor}} > \bar{D}_{\text{TX}}$, we flag the instance as a potential blockage event that will occur within the next $s$ frames, where $s$ was defined in Section \ref{sec:TX_track_bl}.

Following the clearance of a blockage event in the physical environment, the primary objective is to rapidly reestablish the LoS connection to restore reliable communication. Throughout the blockage duration, we maintain continuous execution of the object tracker to monitor all visible objects in the scene. Upon blockage clearance, we propose a two-stage mechanism to re-associate the TX with its original tracker:

\textbf{Stage 1: Predictive Bounding Box Association.} We extrapolate the TX bbox from its last recorded frame prior to occlusion using the motion prediction model defined in \eqref{eqn:pred_bbox}, estimating the expected TX position at the current frame. If an object appears within the vicinity of this predicted bbox, we compute the IoU between the predicted and observed bboxes. If the IoU exceeds a predefined threshold $\tau_{\text{match}}$, we provisionally designate that object as the TX and assign it the original tracker ID maintained before the blockage period.

\textbf{Stage 2: TX Re-identification Verification.} Concurrently with Stage 1, we reinitiate the TX identification procedure detailed in Section \ref{sec:TX_identi_bl} to re-identify the TX from the scene definitively. During the re-identification process, which may require up to $M=3$ consecutive frames for reliable classification (as explained in Section \ref{sec:TX_identi_bl}), we continue tracking the provisionally assigned object from Stage 1. Upon completion of the re-identification process: (i) if the definitively identified TX corresponds to the provisionally tracked object, we maintain the existing tracker association and continue normal operation; (ii) if the provisionally tracked object is incorrect, we reassign the TX designation to the correctly identified object under a new tracker ID and resume the standard blockage prediction workflow. 
This dual-stage recovery strategy ensures robust TX re-acquisition as soon as the blockage clears for efficient LoS communication.

\subsection{Numerical Results}

In this section, we evaluate the effectiveness of our blockage prediction approach. Note that we demonstrate the superiority of our TX identification and beam selection approach earlier in Section \ref{sec:TX_identi_bl} and Section \ref{sec:beam_pred}
As established in Section \ref{sec:va-bi}, this work represents the first systematic investigation of proactive vision-aided blockage prediction for scenarios involving simultaneous non-uniform mobility of both the TX and multiple dynamic obstacles 

The zero-shot prediction capability of our proposed algorithm across diverse conditions demonstrates the robustness and generalizability in novel scenarios. 
Our framework is explicitly designed to prioritize recall over precision, ensuring comprehensive coverage of potential blockage events even at the cost of minimal false positive detections. This design philosophy is motivated by the critical nature of blockage prediction in mmWave systems: failing to predict an actual blockage (false negative) results in complete link failure and service disruption, whereas a false positive prediction triggers unnecessary but non-catastrophic preventive handoff or frequency switching. Therefore, maximizing recall by capturing all genuine blockage instances before they occur takes precedence over minimizing false alarms in our evaluation metrics.

\begin{figure}[t!]
    \centering
    \subfloat[$S = 1$]{
        \includegraphics[width=0.3\linewidth]{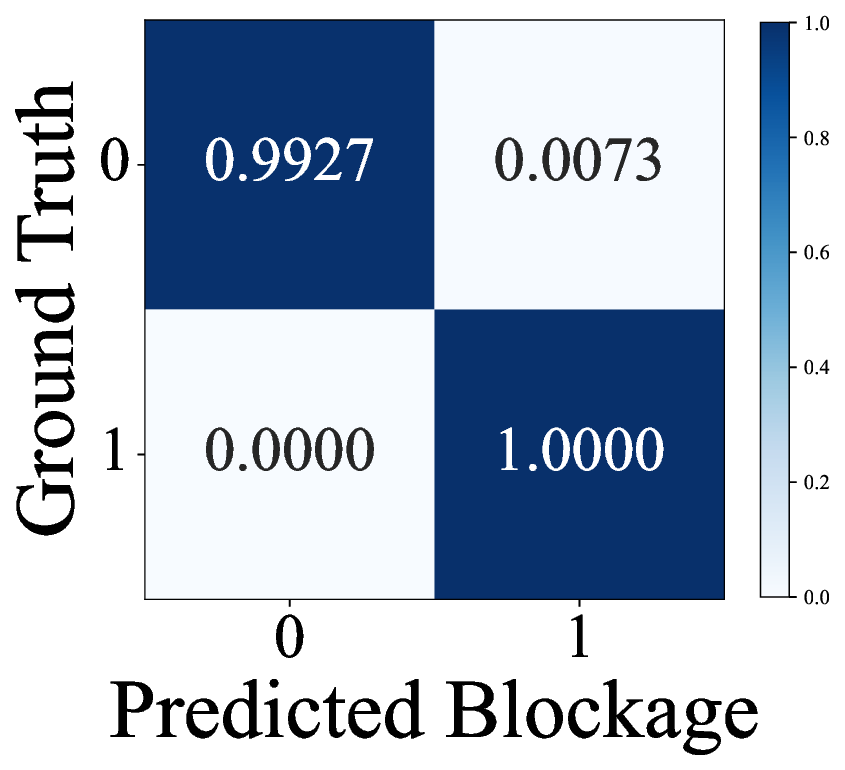}
        \label{fig:s1_10}
    }\hfill
    \subfloat[$S = 2$]{
        \includegraphics[width=0.3\linewidth]{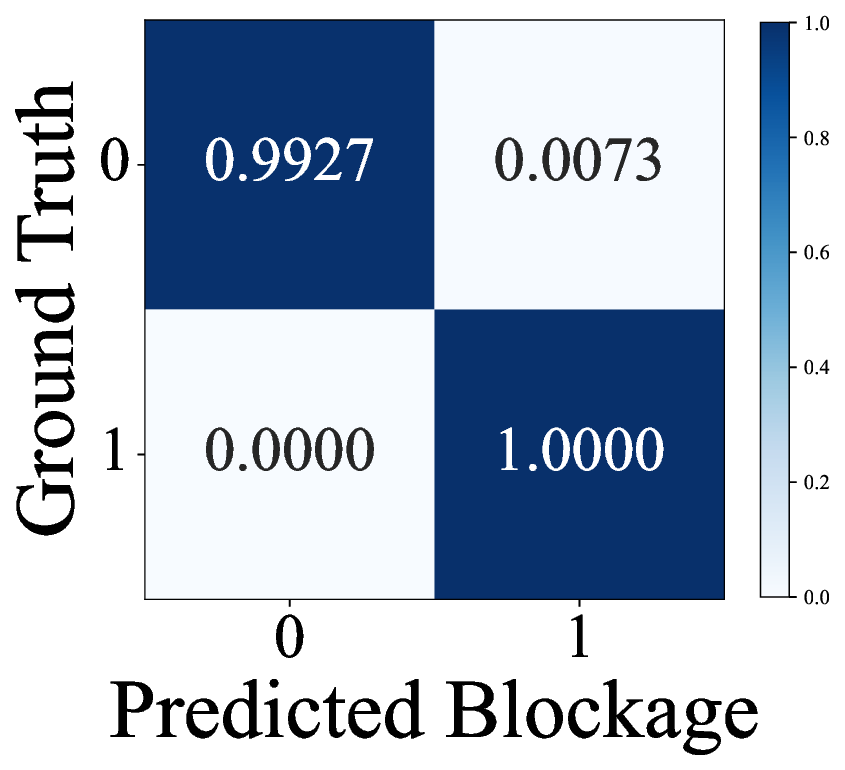}
        \label{fig:s2_10}
    }\hfill
    \subfloat[$S = 3$]{
        \includegraphics[width=0.3\linewidth]{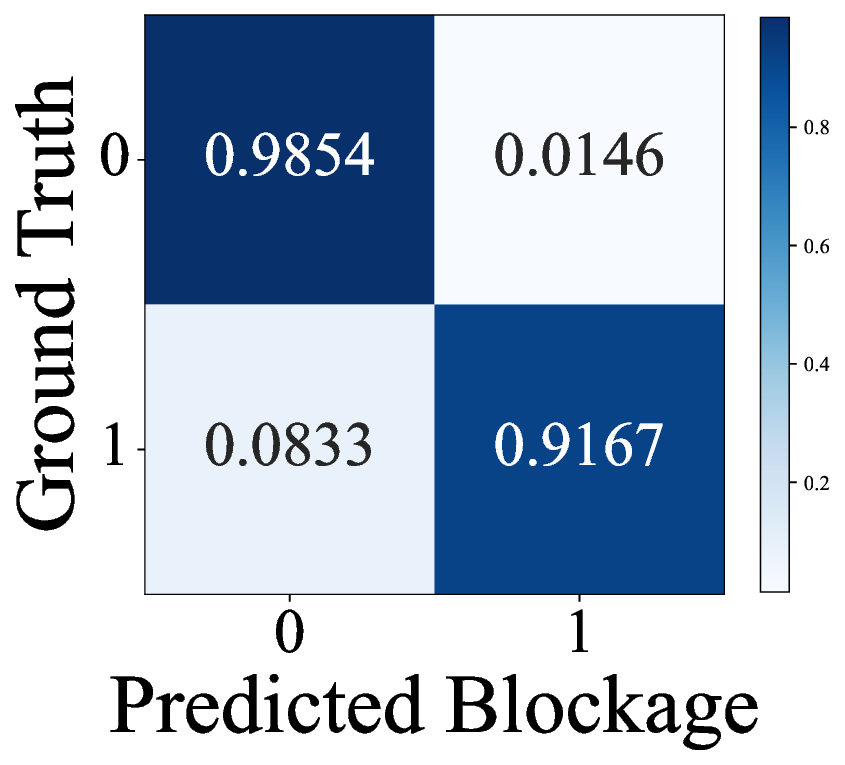}
        \label{fig:s3_10}
    }
    \caption{Ground truth normalized confusion matrices for scenario 10 for future prediction windows $S \in \{1,2,3\}$.}
    \label{fig:scen10}
\end{figure}

\begin{figure}[t!]
    \centering
    \subfloat[$S = 1$]{
        \includegraphics[width=0.3\linewidth]{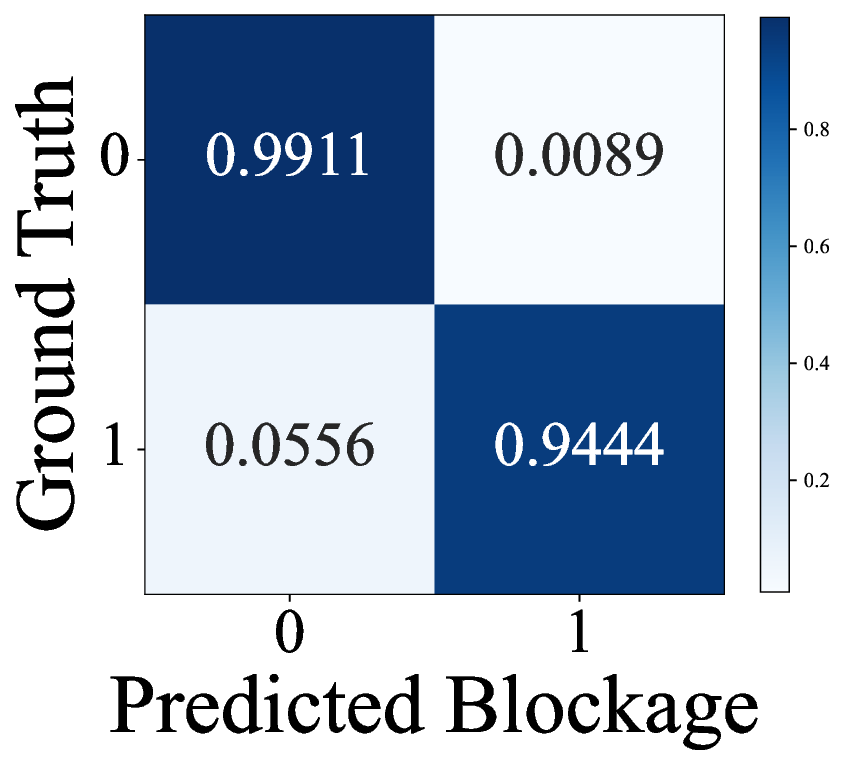}
        \label{fig:s1_11}
    }\hfill
    \subfloat[$S = 2$]{
        \includegraphics[width=0.3\linewidth]{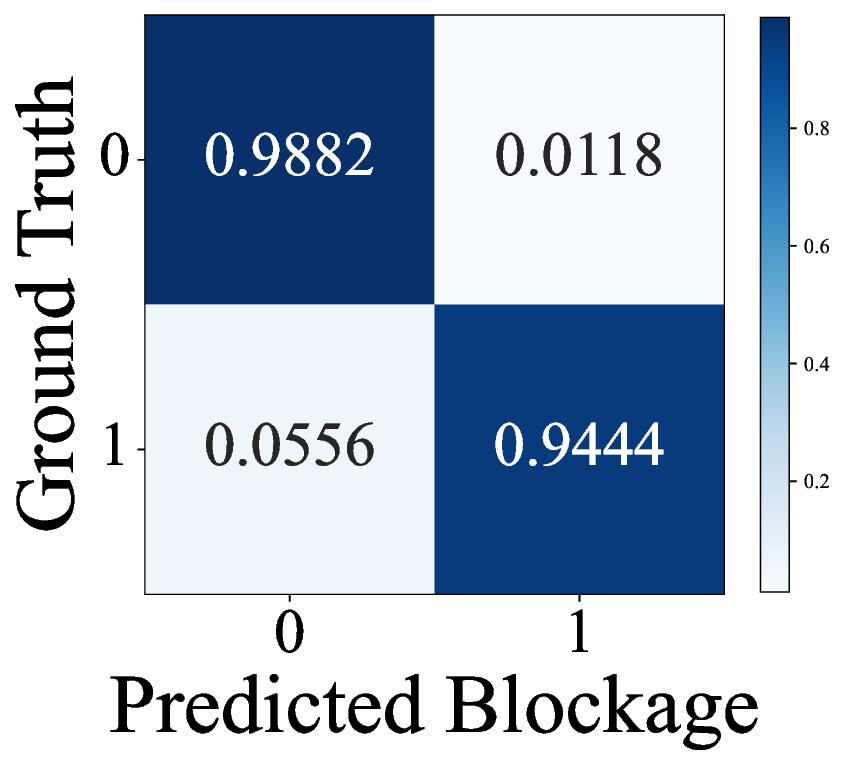}
        \label{fig:s2_11}
    }\hfill
    \subfloat[$S = 3$]{
        \includegraphics[width=0.3\linewidth]{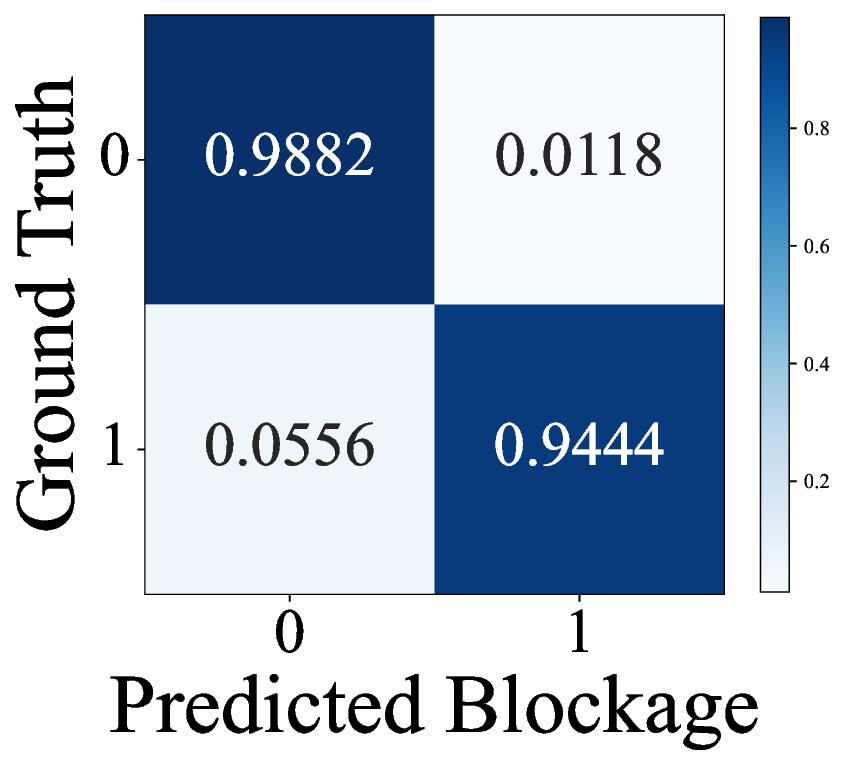}
        \label{fig:s3_11}
    }
    \caption{Ground truth normalized confusion matrices for scenario 11 for future prediction windows $S \in \{1,2,3\}$.}
    \label{fig:scen11}
\end{figure}

\begin{figure}[t!]
    \centering
    \subfloat[$S = 1$]{
        \includegraphics[width=0.3\linewidth]{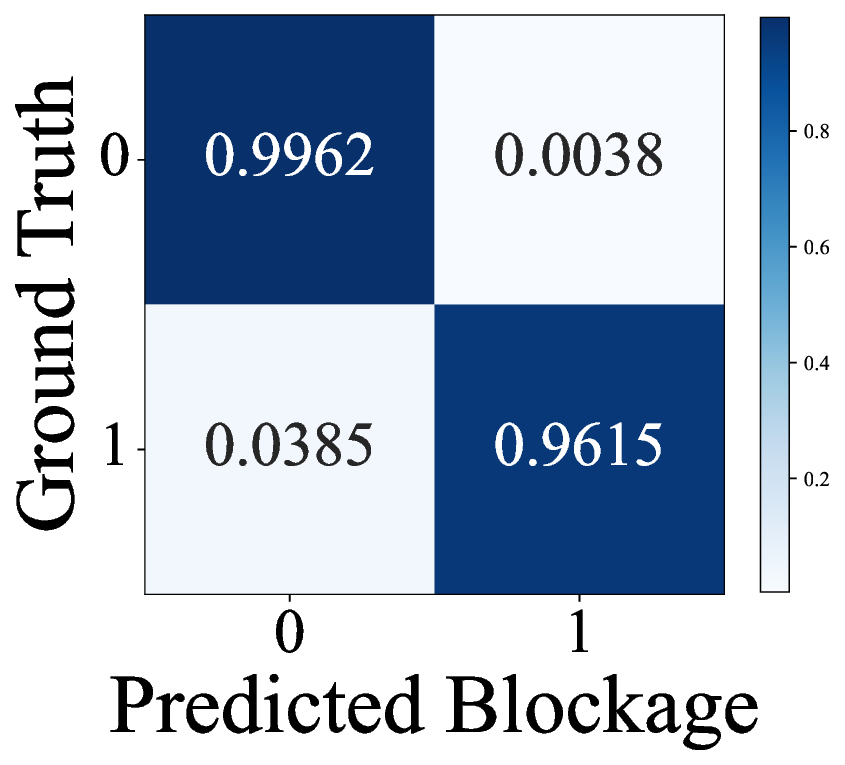}
        \label{fig:s1_12}
    }\hfill
    \subfloat[$S = 2$]{
        \includegraphics[width=0.3\linewidth]{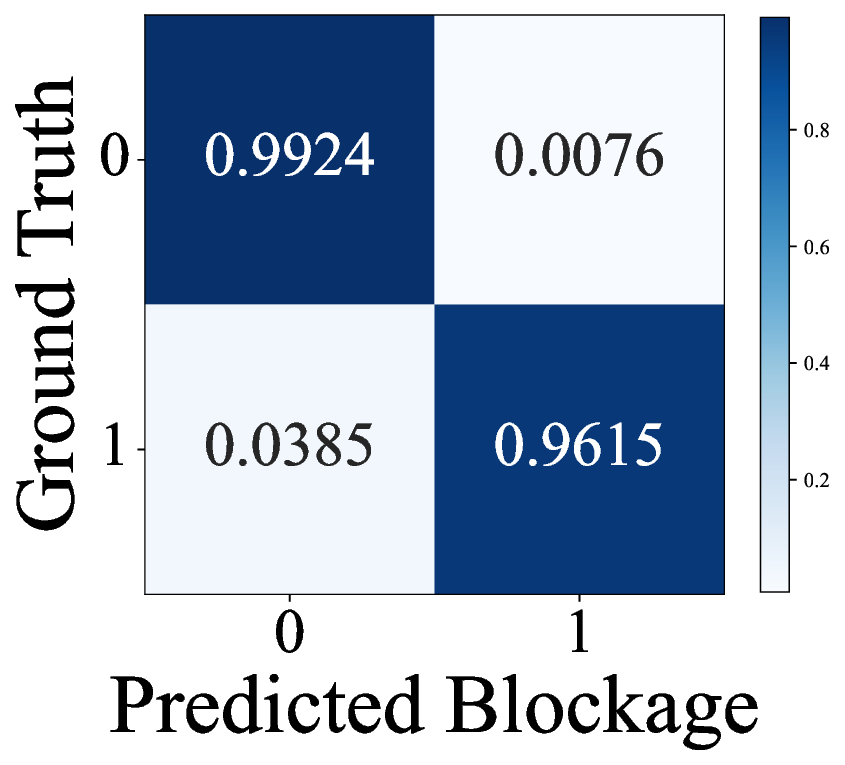}
        \label{fig:s2_12}
    }\hfill
    \subfloat[$S = 3$]{
        \includegraphics[width=0.3\linewidth]{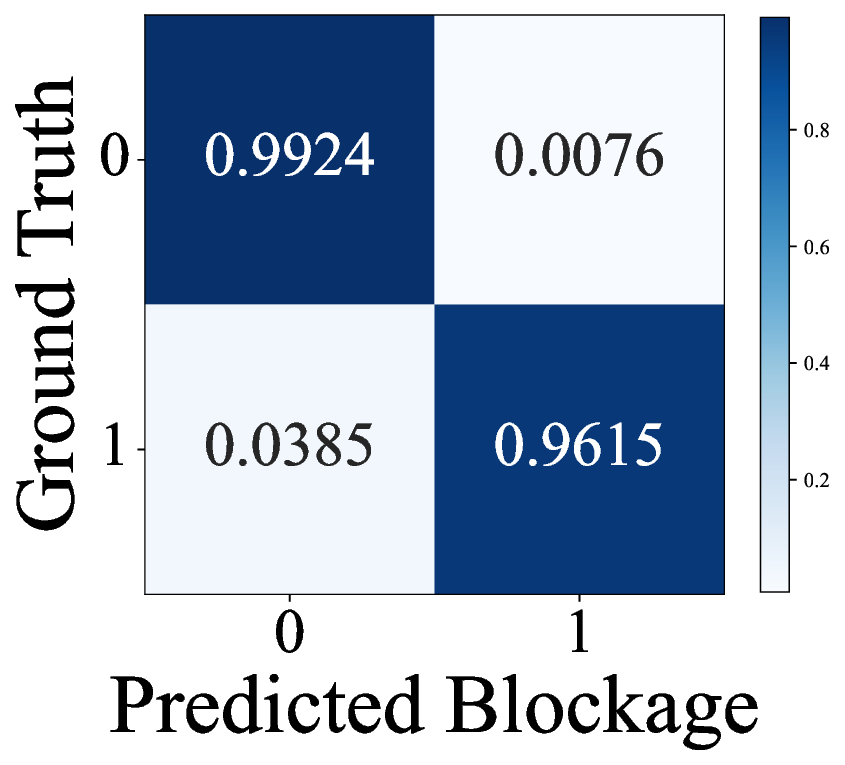}
        \label{fig:s3_12}
    }
    \caption{Ground truth normalized confusion matrices for scenario 12 for future prediction windows $S \in \{1,2,3\}$.}
    \label{fig:scen12}
\end{figure}

Table \ref{tab:blockage_results} presents the blockage prediction accuracy, precision, recall, false positive rate (FPR), and false negative rate (FNR), while Figs. \ref{fig:scen10}, \ref{fig:scen11}, \ref{fig:scen12} represent the ground truth normalized confusion matrices. The effectiveness of our approach is evident by Table \ref{tab:blockage_results}, which shows consistently high performance across scenarios with varying environmental conditions, TX types, and blockage characteristics. We use scenarios 10, 11, and 12 of the DeepSense6G dataset \cite{deepsense6g} to evaluate the proposed framework, which consists of 447, 356, and 289 test samples, respectively.
Our framework predicts blockages up to $S$ frames ahead, where $S \in \{1, 2, 3\}$ as defined in Section \ref{sec:TX_track_bl}. As evident in Table \ref{tab:blockage_results}, the prediction accuracy remains nearly constant across all prediction horizons, achieving near perfect blockage predicting capabilities. This demonstrates the temporal stability and accuracy of our proposed framework. 
The false positive rate increases marginally with longer prediction windows, reflecting a gradual decrease in precision, as shown in Table \ref{tab:blockage_results}. This behavior arises because the inherent uncertainty in trajectory extrapolation accumulates over extended time intervals. The effect is primarily due to the non-uniform movement patterns of both the TX and surrounding distractors.
The recall of our framework remains high and nearly constant even as the prediction window increases, as evident in Figs. \ref{fig:scen10}, \ref{fig:scen11}, \ref{fig:scen12}. This further validates the fundamental goal of our framework: ensuring that genuine blockage instances are reliably flagged while false negatives remain minimal.

\begin{table}[t!]
\centering
\caption{DeepSense 6G Blockage Prediction Accuracy: Models tested on Scenarios 10, 11, and 12. 
}
\label{tab:blockage_results}

\resizebox{0.5\textwidth}{!}{%
\begin{tabular}{c c c c c c c c c}
\toprule
Scenario & \multicolumn{3}{c}{\begin{tabular}[c]{@{}c@{}}Future Prediction \\ Window $(S)$\end{tabular}} & Accuracy & Precision & Recall & FPR & FNR \\ 
\midrule
\multirow{3}{*}{10}
         & \multicolumn{3}{c}{1} & 99.33\% & 0.92 & 1.00  &0.72\%  &0\%  \\
         & \multicolumn{3}{c}{2} & 99.33\% & 0.92 & 1.00 &0.72\%  &0\% \\
         & \multicolumn{3}{c}{3} & 97.99\% & 0.85 & 0.92 &1.46\%  &8.33\%  \\
\midrule
\multirow{3}{*}{11}
         & \multicolumn{3}{c}{1} & 98.87\% & 0.85 & 0.94 & 0.89\%  & 5.55\%\\
         & \multicolumn{3}{c}{2} & 98.60\% & 0.81 & 0.94 & 1.18\%  & 5.55\% \\
         & \multicolumn{3}{c}{3} & 98.60\% & 0.81 & 0.94 & 1.18\% & 5.55\%  \\
\midrule
\multirow{3}{*}{12}
         & \multicolumn{3}{c}{1} & 99.31\% & 0.96 & 0.96 & 0.38\% & 3.85\% \\
         & \multicolumn{3}{c}{2} & 98.96\% & 0.93 & 0.96 & 0.76\% & 3.85\% \\
         & \multicolumn{3}{c}{3} & 98.96\% & 0.93 & 0.96 & 0.76\% & 3.85\% \\
\bottomrule
\end{tabular}
}
\end{table}


\section{Conclusion}
This paper presents two distinct frameworks addressing the fundamental challenges in mmWave communications, namely, propagation loss and penetration loss. By integrating RGB imagery as auxiliary sensory input, both frameworks eliminate the need for computationally expensive exhaustive beam sweeping.
Our propagation loss mitigation framework surpasses current vision-aided state-of-the-art methods by at least 6\% across Top-1, Top-3, and Top-5 beam prediction metrics, demonstrating superior performance with reduced computational overhead.
Our penetration loss mitigation framework represents the first investigation of blockage prediction for scenarios with simultaneous dynamic mobility of both the TX and multiple obstacles. 
This framework achieves over 98\% accuracy in predicting blockages up to three frames ahead, establishing a strong benchmark for this previously unexplored problem and demonstrating that vision-aided approaches can effectively enable proactive beam management in highly dynamic environments.
Future extensions of this work include predicting self-blockages and using the TX location for vision-aided BS handover. Predicted TX positions during blockages can enable coordinated handovers to alternative BSs equipped with cameras to ensure seamless connectivity.

\bibliographystyle{IEEEtran}
\bibliography{mybib}

\end{document}